\ifx\newheadisloaded\relax\immediate\write16{***already loaded}\endinput\else\let\newheadisloaded=\relax\fi
\gdef\isonnarrowscreen{F}
\gdef\PSfonts{T}
\magnification\magstep1

\newdimen\papwidth
\newdimen\papheight
\newskip\beforesectionskipamount  
\newskip\sectionskipamount 
\def\sectionskip{\vskip\sectionskipamount}
\def\beforesectionskip{\vskip\beforesectionskipamount}
\papwidth=16truecm
\if F\isonnarrowscreen
\papheight=22truecm
\voffset=0.4truecm
\hoffset=0.4truecm
\else
\papheight=22truecm
\voffset=-1.5truecm
\hoffset=-1truecm
\fi
\hsize=\papwidth
\vsize=\papheight
\catcode`\@=11
\ifx\amstexloaded@\relax
\else
\nopagenumbers
\headline={\ifnum\pageno>1 {\hss\tenrm-\ \folio\ -\hss} \else
{\hfill}\fi}
\fi
\catcode`\@=\active
\newdimen\texpscorrection
\texpscorrection=0.15truecm 

\def\sectionsize{\twelvepoint}
\def\sectiontype{\bf}
\def\subsectionsize{}
\def\subsectiontype{\bf}
\def\em{\sl}
\newfam\truecmsy
\newfam\truecmr
\newfam\msbfam
\newfam\scriptfam
\newfam\frakfam
\newfam\frakbfam

\newskip\ttglue 
\if T\isonnarrowscreen
\papheight=11.5truecm
\fi
\if F\PSfonts
\font\twelverm=cmr12
\font\tenrm=cmr10
\font\eightrm=cmr8
\font\sevenrm=cmr7
\font\sixrm=cmr6
\font\fiverm=cmr5

\font\twelvebf=cmbx12
\font\tenbf=cmbx10
\font\eightbf=cmbx8
\font\sevenbf=cmbx7
\font\sixbf=cmbx6
\font\fivebf=cmbx5

\font\twelveit=cmti12
\font\tenit=cmti10
\font\eightit=cmti8
\font\sevenit=cmti7
\font\sixit=cmti6
\font\fiveit=cmti5

\font\twelvesl=cmsl12
\font\tensl=cmsl10
\font\eightsl=cmsl8
\font\sevensl=cmsl7
\font\sixsl=cmsl6
\font\fivesl=cmsl5

\font\twelvei=cmmi12
\font\teni=cmmi10
\font\eighti=cmmi8
\font\seveni=cmmi7
\font\sixi=cmmi6
\font\fivei=cmmi5

\font\twelvesy=cmsy10	at	12pt
\font\tensy=cmsy10
\font\eightsy=cmsy8
\font\sevensy=cmsy7
\font\sixsy=cmsy6
\font\fivesy=cmsy5
\font\twelvetruecmsy=cmsy10	at	12pt
\font\tentruecmsy=cmsy10
\font\eighttruecmsy=cmsy8
\font\seventruecmsy=cmsy7
\font\sixtruecmsy=cmsy6
\font\fivetruecmsy=cmsy5

\font\twelvetruecmr=cmr12
\font\tentruecmr=cmr10
\font\eighttruecmr=cmr8
\font\seventruecmr=cmr7
\font\sixtruecmr=cmr6
\font\fivetruecmr=cmr5

\font\twelvebf=cmbx12
\font\tenbf=cmbx10
\font\eightbf=cmbx8
\font\sevenbf=cmbx7
\font\sixbf=cmbx6
\font\fivebf=cmbx5

\font\twelvett=cmtt12
\font\tentt=cmtt10
\font\eighttt=cmtt8

\font\twelveex=cmex10	at	12pt
\font\tenex=cmex10

\font\twelvemsb=msbm10	at	12pt
\font\tenmsb=msbm10
\font\eightmsb=msbm8
\font\sevenmsb=msbm7
\font\sixmsb=msbm6
\font\fivemsb=msbm5


\font\tenfrm=eufm10
\font\eightfrm=eufm8
\font\sevenfrm=eufm7
\font\sixfrm=eufm6
\font\fivefrm=eufm5

\font\tenfrb=eufb10
\font\eightfrb=eufb8
\font\sevenfrb=eufb7
\font\sixfrb=eufb6
\font\fivefrb=eufb5
\font\twelvescr=eusm10 at 12pt
\font\tenscr=eusm10
\font\eightscr=eusm8
\font\sevenscr=eusm7
\font\sixscr=eusm6
\font\fivescr=eusm5
\fi
\if T\PSfonts
\font\twelverm=ptmr	at	12pt
\font\tenrm=ptmr	at	10pt
\font\eightrm=ptmr	at	8pt
\font\sevenrm=ptmr	at	7pt
\font\sixrm=ptmr	at	6pt
\font\fiverm=ptmr	at	5pt

\font\twelvebf=ptmb	at	12pt
\font\tenbf=ptmb	at	10pt
\font\eightbf=ptmb	at	8pt
\font\sevenbf=ptmb	at	7pt
\font\sixbf=ptmb	at	6pt
\font\fivebf=ptmb	at	5pt

\font\twelveit=ptmri	at	12pt
\font\tenit=ptmri	at	10pt
\font\eightit=ptmri	at	8pt
\font\sevenit=ptmri	at	7pt
\font\sixit=ptmri	at	6pt
\font\fiveit=ptmri	at	5pt

\font\twelvesl=ptmro	at	12pt
\font\tensl=ptmro	at	10pt
\font\eightsl=ptmro	at	8pt
\font\sevensl=ptmro	at	7pt
\font\sixsl=ptmro	at	6pt
\font\fivesl=ptmro	at	5pt

\font\twelvei=cmmi12
\font\teni=cmmi10
\font\eighti=cmmi8
\font\seveni=cmmi7
\font\sixi=cmmi6
\font\fivei=cmmi5

\font\twelvesy=cmsy10	at	12pt
\font\tensy=cmsy10
\font\eightsy=cmsy8
\font\sevensy=cmsy7
\font\sixsy=cmsy6
\font\fivesy=cmsy5
\font\twelvetruecmsy=cmsy10	at	12pt
\font\tentruecmsy=cmsy10
\font\eighttruecmsy=cmsy8
\font\seventruecmsy=cmsy7
\font\sixtruecmsy=cmsy6
\font\fivetruecmsy=cmsy5

\font\twelvetruecmr=cmr12
\font\tentruecmr=cmr10
\font\eighttruecmr=cmr8
\font\seventruecmr=cmr7
\font\sixtruecmr=cmr6
\font\fivetruecmr=cmr5


\font\twelvett=cmtt12
\font\tentt=cmtt10
\font\eighttt=cmtt8

\font\twelveex=cmex10	at	12pt
\font\tenex=cmex10

\font\twelvemsb=msbm10	at	12pt
\font\tenmsb=msbm10
\font\eightmsb=msbm8
\font\sevenmsb=msbm7
\font\sixmsb=msbm6
\font\fivemsb=msbm5


\font\tenfrm=eufm10
\font\eightfrm=eufm8
\font\sevenfrm=eufm7
\font\sixfrm=eufm6
\font\fivefrm=eufm5

\font\tenfrb=eufb10
\font\eightfrb=eufb8
\font\sevenfrb=eufb7
\font\sixfrb=eufb6
\font\fivefrb=eufb5
\font\twelvescr=eusm10 at 12pt
\font\tenscr=eusm10
\font\eightscr=eusm8
\font\sevenscr=eusm7
\font\sixscr=eusm6
\font\fivescr=eusm5
\fi
\def\eightpoint{\def\rm{\fam0\eightrm}%
\textfont0=\eightrm
  \scriptfont0=\sixrm
  \scriptscriptfont0=\fiverm 
\textfont1=\eighti
  \scriptfont1=\sixi
  \scriptscriptfont1=\fivei 
\textfont2=\eightsy
  \scriptfont2=\sixsy
  \scriptscriptfont2=\fivesy 
\textfont3=\tenex
  \scriptfont3=\tenex
  \scriptscriptfont3=\tenex 
\textfont\itfam=\eightit
  \scriptfont\itfam=\sixit
  \scriptscriptfont\itfam=\fiveit 
  \def\it{\fam\itfam\eightit}%
\textfont\slfam=\eightsl
  \scriptfont\slfam=\sixsl
  \scriptscriptfont\slfam=\fivesl 
  \def\sl{\fam\slfam\eightsl}%
\textfont\ttfam=\eighttt
  \def\tt{\fam\ttfam\eighttt}%
\textfont\bffam=\eightbf
  \scriptfont\bffam=\sixbf
  \scriptscriptfont\bffam=\fivebf
  \def\bf{\fam\bffam\eightbf}%
\textfont\frakfam=\eightfrm
  \scriptfont\frakfam=\sixfrm
  \scriptscriptfont\frakfam=\fivefrm
  \def\frak{\fam\frakfam\eightfrm}%
\textfont\frakbfam=\eightfrb
  \scriptfont\frakbfam=\sixfrb
  \scriptscriptfont\frakbfam=\fivefrb
  \def\bfrak{\fam\frakbfam\eightfrb}%
\textfont\scriptfam=\eightscr
  \scriptfont\scriptfam=\sixscr
  \scriptscriptfont\scriptfam=\fivescr
  \def\script{\fam\scriptfam\eightscr}%
\textfont\msbfam=\eightmsb
  \scriptfont\msbfam=\sixmsb
  \scriptscriptfont\msbfam=\fivemsb
  \def\bb{\fam\msbfam\eightmsb}%
\textfont\truecmr=\eighttruecmr
  \scriptfont\truecmr=\sixtruecmr
  \scriptscriptfont\truecmr=\fivetruecmr
  \def\truerm{\fam\truecmr\eighttruecmr}%
\textfont\truecmsy=\eighttruecmsy
  \scriptfont\truecmsy=\sixtruecmsy
  \scriptscriptfont\truecmsy=\fivetruecmsy
\tt \ttglue=.5em plus.25em minus.15em 
\normalbaselineskip=9pt
\setbox\strutbox=\hbox{\vrule height7pt depth2pt width0pt}%
\normalbaselines
\rm
}

\def\tenpoint{\def\rm{\fam0\tenrm}%
\textfont0=\tenrm
  \scriptfont0=\sevenrm
  \scriptscriptfont0=\fiverm 
\textfont1=\teni
  \scriptfont1=\seveni
  \scriptscriptfont1=\fivei 
\textfont2=\tensy
  \scriptfont2=\sevensy
  \scriptscriptfont2=\fivesy 
\textfont3=\tenex
  \scriptfont3=\tenex
  \scriptscriptfont3=\tenex 
\textfont\itfam=\tenit
  \scriptfont\itfam=\sevenit
  \scriptscriptfont\itfam=\fiveit 
  \def\it{\fam\itfam\tenit}%
\textfont\slfam=\tensl
  \scriptfont\slfam=\sevensl
  \scriptscriptfont\slfam=\fivesl 
  \def\sl{\fam\slfam\tensl}%
\textfont\ttfam=\tentt
  \def\tt{\fam\ttfam\tentt}%
\textfont\bffam=\tenbf
  \scriptfont\bffam=\sevenbf
  \scriptscriptfont\bffam=\fivebf
  \def\bf{\fam\bffam\tenbf}%
\textfont\frakfam=\tenfrm
  \scriptfont\frakfam=\sevenfrm
  \scriptscriptfont\frakfam=\fivefrm
  \def\frak{\fam\frakfam\tenfrm}%
\textfont\frakbfam=\tenfrb
  \scriptfont\frakbfam=\sevenfrb
  \scriptscriptfont\frakbfam=\fivefrb
  \def\bfrak{\fam\frakbfam\tenfrb}%
\textfont\scriptfam=\tenscr
  \scriptfont\scriptfam=\sevenscr
  \scriptscriptfont\scriptfam=\fivescr
  \def\script{\fam\scriptfam\tenscr}%
\textfont\msbfam=\tenmsb
  \scriptfont\msbfam=\sevenmsb
  \scriptscriptfont\msbfam=\fivemsb
  \def\bb{\fam\msbfam\tenmsb}%
\textfont\truecmr=\tentruecmr
  \scriptfont\truecmr=\seventruecmr
  \scriptscriptfont\truecmr=\fivetruecmr
  \def\truerm{\fam\truecmr\tentruecmr}%
\textfont\truecmsy=\tentruecmsy
  \scriptfont\truecmsy=\seventruecmsy
  \scriptscriptfont\truecmsy=\fivetruecmsy
\tt \ttglue=.5em plus.25em minus.15em 
\normalbaselineskip=12pt
\setbox\strutbox=\hbox{\vrule height8.5pt depth3.5pt width0pt}%
\normalbaselines
\rm
}

\def\twelvepoint{\def\rm{\fam0\twelverm}%
\textfont0=\twelverm
  \scriptfont0=\tenrm
  \scriptscriptfont0=\eightrm 
\textfont1=\twelvei
  \scriptfont1=\teni
  \scriptscriptfont1=\eighti 
\textfont2=\twelvesy
  \scriptfont2=\tensy
  \scriptscriptfont2=\eightsy 
\textfont3=\twelveex
  \scriptfont3=\twelveex
  \scriptscriptfont3=\twelveex 
\textfont\itfam=\twelveit
  \scriptfont\itfam=\tenit
  \scriptscriptfont\itfam=\eightit 
  \def\it{\fam\itfam\twelveit}%
\textfont\slfam=\twelvesl
  \scriptfont\slfam=\tensl
  \scriptscriptfont\slfam=\eightsl 
  \def\sl{\fam\slfam\twelvesl}%
\textfont\ttfam=\twelvett
  \def\tt{\fam\ttfam\twelvett}%
\textfont\bffam=\twelvebf
  \scriptfont\bffam=\tenbf
  \scriptscriptfont\bffam=\eightbf
  \def\bf{\fam\bffam\twelvebf}%
\textfont\scriptfam=\twelvescr
  \scriptfont\scriptfam=\tenscr
  \scriptscriptfont\scriptfam=\eightscr
  \def\script{\fam\scriptfam\twelvescr}%
\textfont\msbfam=\twelvemsb
  \scriptfont\msbfam=\tenmsb
  \scriptscriptfont\msbfam=\eightmsb
  \def\bb{\fam\msbfam\twelvemsb}%
\textfont\truecmr=\twelvetruecmr
  \scriptfont\truecmr=\tentruecmr
  \scriptscriptfont\truecmr=\eighttruecmr
  \def\truerm{\fam\truecmr\twelvetruecmr}%
\textfont\truecmsy=\twelvetruecmsy
  \scriptfont\truecmsy=\tentruecmsy
  \scriptscriptfont\truecmsy=\eighttruecmsy
\tt \ttglue=.5em plus.25em minus.15em 
\setbox\strutbox=\hbox{\vrule height7pt depth2pt width0pt}%
\normalbaselineskip=15pt
\normalbaselines
\rm
}
%
\fontdimen16\tensy=2.7pt
\fontdimen13\tensy=4.3pt
\fontdimen17\tensy=2.7pt
\fontdimen14\tensy=4.3pt
\fontdimen18\tensy=4.3pt
\fontdimen16\eightsy=2.7pt
\fontdimen13\eightsy=4.3pt
\fontdimen17\eightsy=2.7pt
\fontdimen14\eightsy=4.3pt
\fontdimen18\sevensy=4.3pt
\fontdimen16\sevensy=1.8pt
\fontdimen13\sevensy=4.3pt
\fontdimen17\sevensy=2.7pt
\fontdimen14\sevensy=4.3pt
\fontdimen18\sevensy=4.3pt
%
\def\hexnumber#1{\ifcase#1 0\or1\or2\or3\or4\or5\or6\or7\or8\or9\or
 A\or B\or C\or D\or E\or F\fi}
\mathcode`\=="3\hexnumber\truecmr3D
\mathchardef\not="3\hexnumber\truecmsy36
\mathcode`\+="2\hexnumber\truecmr2B
\mathcode`\(="4\hexnumber\truecmr28
\mathcode`\)="5\hexnumber\truecmr29
\mathcode`\!="5\hexnumber\truecmr21
\mathcode`\(="4\hexnumber\truecmr28
\mathcode`\)="5\hexnumber\truecmr29

\def\tilde{\mathaccent"0\hexnumber\truecmr7E }
\def\bar{\mathaccent"0\hexnumber\truecmr16 }

\def\hat{\mathaccent"0\hexnumber\truecmr5E }

\def\Phi{\mathchar"0\hexnumber\truecmr08 }
\def\Gamma {\mathchar"0\hexnumber\truecmr00 }
\def\Delta {\mathchar"0\hexnumber\truecmr01 }
\def\Theta {\mathchar"0\hexnumber\truecmr02 }
\def\Lambda{\mathchar"0\hexnumber\truecmr03 }
\def\Xi {\mathchar"0\hexnumber\truecmr04 }
\def\Pi{\mathchar"0\hexnumber\truecmr05 }
\def\Sigma{\mathchar"0\hexnumber\truecmr06 }
\def\Upsilon {\mathchar"0\hexnumber\truecmr07 }
\def\Phi {\mathchar"0\hexnumber\truecmr08 }
\def\Psi {\mathchar"0\hexnumber\truecmr09 }
\def\Omega{\mathchar"0\hexnumber\truecmr0A }
\newcount\EQNcount \EQNcount=1
\newcount\CLAIMcount \CLAIMcount=1
\newcount\SECTIONcount \SECTIONcount=0
\newcount\SUBSECTIONcount \SUBSECTIONcount=1
\def\ifff#1#2#3{\ifundefined{#1#2}%
\expandafter\xdef\csname #1#2\endcsname{#3}\else%
\immediate\write16{!!!!!doubly defined #1,#2}\fi}
\def\NEWDEF#1#2#3{\ifff{#1}{#2}{#3}}
\def\actualnumber{\number\SECTIONcount}
\def\EQ#1{\lmargin{#1}\eqno\tageck{#1}}
\def\NR#1{&\lmargin{#1}\tageck{#1}\cr}  
\def\tageck#1{\lmargin{#1}({\rm \actualnumber}.\number\EQNcount)
 \NEWDEF{e}{#1}{(\actualnumber.\number\EQNcount)}
\global\advance\EQNcount by 1
}
\def\SECT#1#2{\lmargin{#1}\SECTION{#2}%
\NEWDEF{s}{#1}{\actualnumber}%
}
\def\SUBSECT#1#2{\lmargin{#1}
\NEWDEF{s}{#1}{\actualnumber.\number\SUBSECTIONcount}%
\SUBSECTION{#2}%
}
\def\CLAIM#1#2#3\par{
\vskip.1in\medbreak\noindent
{\lmargin{#2}\bf #1\ \actualnumber.\number\CLAIMcount.} {\sl #3}\par
\NEWDEF{c}{#2}{#1\ \actualnumber.\number\CLAIMcount}
\global\advance\CLAIMcount by 1
\ifdim\lastskip<\medskipamount
\removelastskip\penalty55\medskip\fi}
\def\CLAIMNONR #1#2#3\par{
\vskip.1in\medbreak\noindent
{\lmargin{#2}\bf #1.} {\sl #3}\par
\NEWDEF{c}{#2}{#1}
\global\advance\CLAIMcount by 1
\ifdim\lastskip<\medskipamount
\removelastskip\penalty55\medskip\fi}
\def\SECTION#1{\vskip0pt plus.3\vsize\penalty-75
    \vskip0pt plus -.3\vsize
    \global\advance\SECTIONcount by 1
    \beforesectionskip\noindent
{\sectionsize\sectiontype \actualnumber.\ #1}
    \EQNcount=1
    \CLAIMcount=1
    \SUBSECTIONcount=1
    \nobreak\sectionskip\noindent}
\def\SECTIONNONR#1{\vskip0pt plus.3\vsize\penalty-75
    \vskip0pt plus -.3\vsize
    \global\advance\SECTIONcount by 1
    \beforesectionskip\noindent
{\sectionsize\sectiontype  #1}
     \EQNcount=1
     \CLAIMcount=1
     \SUBSECTIONcount=1
     \nobreak\sectionskip\noindent}
\def\SUBSECTION#1{\vskip0pt plus.2\vsize\penalty-75%
    \vskip0pt plus -.2\vsize%
    \beforesectionskip\noindent%
{\subsectionsize\subsectiontype \actualnumber.\number\SUBSECTIONcount.\ #1}
    \global\advance\SUBSECTIONcount by 1
    \nobreak\sectionskip\noindent}
\def\SUBSECTIONNONR#1\par{\vskip0pt plus.2\vsize\penalty-75
    \vskip0pt plus -.2\vsize
\beforesectionskip\noindent
{\subsectionsize\subsectiontype #1}
    \nobreak\sectionskip\noindent\noindent}
\def\ifundefined#1{\expandafter\ifx\csname#1\endcsname\relax}
\def\equ#1{\ifundefined{e#1}$\spadesuit$#1\else\csname e#1\endcsname\fi}
\def\clm#1{\ifundefined{c#1}$\spadesuit$#1\else\csname c#1\endcsname\fi}
\def\sec#1{\ifundefined{s#1}$\spadesuit$#1
\else Section \csname s#1\endcsname\fi}
\def\lab#1#2{\ifundefined{#1#2}$\spadesuit$#2\else\csname #1#2\endcsname\fi}
\def\fig#1{\ifundefined{fig#1}$\spadesuit$#1\else\csname fig#1\endcsname\fi}
\let\endarg=\par
\def\finish{\def\endarg{\par\endgroup}}
\def\start{\endarg\begingroup}

 \def\beginFROM{\start\parskip=0pt\vskip\baselineskip
\def\finish{\def\endarg{\egroup\par\endgroup}}
  \vbox\bgroup\obeylines\eightpoint\em\finish}

\def\ABSTRACT#1\par{
\vskip 1in {\noindent\sectionsize\sectiontype Abstract.} #1 \par}

\def\TODAY{\number\day~\ifcase\month\or January \or February \or March \or
April \or May \or June
\or July \or August \or September \or October \or November \or December \fi
\number\year\timecount=\number\time
\divide\timecount by 60
}
\newcount\timecount
\def\DRAFT{\def\lmargin##1{\strut\vadjust{\kern-\strutdepth
\vtop to \strutdepth{
\baselineskip\strutdepth\vss\rlap{\kern-1.2 truecm\eightpoint{##1}}}}}
\font\footfont=cmti7
\footline={{\footfont \hfil File:\jobname, \TODAY,  \number\timecount h}}
}
\newbox\strutboxJPE
\setbox\strutboxJPE=\hbox{\strut}
\def\subitem#1#2\par{\vskip\baselineskip\vskip-\ht\strutboxJPE{\item{#1}#2}}
\gdef\strutdepth{\dp\strutbox}
\def\lmargin#1{}
\def\hexnumber#1{\ifcase#1 0\or1\or2\or3\or4\or5\or6\or7\or8\or9\or
 A\or B\or C\or D\or E\or F\fi}
\textfont\msbfam=\tenmsb
\scriptfont\msbfam=\sevenmsb
\scriptscriptfont\msbfam=\fivemsb
\mathchardef\varkappa="0\hexnumber\msbfam7B%
\newcount\FIGUREcount \FIGUREcount=0
\newdimen\figcenter
\def\definefigure#1{\global\advance\FIGUREcount by 1%
\NEWDEF{fig}{#1}{Fig.\ \number\FIGUREcount}
\immediate\write16{  FIG \number\FIGUREcount : #1}}
\def\figure#1#2#3#4\cr{\null%
\definefigure{#1}
{\goodbreak\figcenter=\hsize\relax
\advance\figcenter by -#3truecm
\divide\figcenter by 2
\midinsert\vskip #2truecm\noindent\hskip\figcenter
\includegraphics{#1}\vskip 0.8truecm\noindent \vbox{\eightpoint\noindent
{\bf\fig{#1}}: #4}\endinsert}}
\def\figurewithtex#1#2#3#4#5\cr{\null%
\definefigure{#1}
{\goodbreak\figcenter=\hsize\relax
\advance\figcenter by -#4truecm
\divide\figcenter by 2
\midinsert\vskip #3truecm\noindent\hskip\figcenter
\includegraphics{#1}{\hskip\texpscorrection\input #2 }\vskip 0.8truecm\noindent \vbox{\eightpoint\noindent
{\bf\fig{#1}}: #5}\endinsert}}
\def\figurewithtexplus#1#2#3#4#5#6\cr{\null%
\definefigure{#1}
{\goodbreak\figcenter=\hsize\relax
\advance\figcenter by -#4truecm
\divide\figcenter by 2
\midinsert\vskip #3truecm\noindent\hskip\figcenter
\includegraphics{#1}{\hskip\texpscorrection\input #2 }\vskip #5truecm\noindent \vbox{\eightpoint\noindent
{\bf\fig{#1}}: #6}\endinsert}}
\catcode`@=11
\def\footnote#1{\let\@sf\empty 
  \ifhmode\edef\@sf{\spacefactor\the\spacefactor}\/\fi
  #1\@sf\vfootnote{#1}}
\def\vfootnote#1{\insert\footins\bgroup\eightpoint
  \interlinepenalty\interfootnotelinepenalty
  \splittopskip\ht\strutbox 
  \splitmaxdepth\dp\strutbox \floatingpenalty\@MM
  \leftskip\z@skip \rightskip\z@skip \spaceskip\z@skip \xspaceskip\z@skip
  \textindent{#1}\footstrut\futurelet\next\fo@t}
\def\fo@t{\ifcat\bgroup\noexpand\next \let\next\f@@t
  \else\let\next\f@t\fi \next}
\def\f@@t{\bgroup\aftergroup\@foot\let\next}
\def\f@t#1{#1\@foot}
\def\@foot{\strut\egroup}
\def\footstrut{\vbox to\splittopskip{}}
\skip\footins=\bigskipamount 
\count\footins=1000 
\dimen\footins=8in 
\catcode`@=12 

\def\AA{{\script A}}

\def\CC{{\script C}}

\def\LL{{\script L}}
\def\MM{{\script M}}
\def\NN{{\script N}}
\def\OO{{\script O}}

\def\TT{{\script T}}

\def\HALF{{\textstyle{1\over 2}}}

\def\QEDD{\hfill\smallskip
         \line{$\hfill{\vcenter{\vbox{\hrule height 0.2pt
	\hbox{\vrule width 0.2pt height 1.3ex \kern 1.3ex
		\vrule width 0.2pt}
	\hrule height 0.2pt}}}$
               \ \ \ \ \ \ }
         \bigskip}
\def\QED{$\hfill{\vcenter{\vbox{\hrule height 0.2pt
	\hbox{\vrule width 0.2pt height 1.3ex \kern 1.3ex
		\vrule width 0.2pt}
	\hrule height 0.2pt}}}$\bigskip}
\def\real{{\bf R}}
\def\natural{{\bf N}}
\def\complex{{\bf C}}
\def\integer{{\bf Z}}

\def\Im{{\rm Im\,}}
\def\PROOF{\medskip\noindent{\bf Proof.\ }}
\def\REMARK{\medskip\noindent{\bf Remark.\ }}
\def\LIKEREMARK#1{\medskip\noindent{\bf #1.\ }}
\normalbaselineskip=5.25mm
\baselineskip=5.25mm
\parskip=10pt
\beforesectionskipamount=24pt plus8pt minus8pt
\sectionskipamount=3pt plus1pt minus1pt
\def\em{\it}
\tenpoint
\null
\catcode`\@=11
\ifx\amstexloaded@\relax\catcode`\@=\active
\endinput\fi
\catcode`\@=\active
\def\period{\unskip.\spacefactor3000 { }}
%
%
\newbox\noboxJPE
\newbox\byboxJPE
\newbox\paperboxJPE
\newbox\yrboxJPE
\newbox\jourboxJPE
\newbox\pagesboxJPE
\newbox\volboxJPE
\newbox\preprintboxJPE
\newbox\toappearboxJPE
\newbox\bookboxJPE
\newbox\bybookboxJPE
\newbox\publisherboxJPE
\newbox\inprintboxJPE
\def\refclearJPE{
   \setbox\noboxJPE=\null             \gdef\isnoJPE{F}
   \setbox\byboxJPE=\null             \gdef\isbyJPE{F}
   \setbox\paperboxJPE=\null          \gdef\ispaperJPE{F}
   \setbox\yrboxJPE=\null             \gdef\isyrJPE{F}
   \setbox\jourboxJPE=\null           \gdef\isjourJPE{F}
   \setbox\pagesboxJPE=\null          \gdef\ispagesJPE{F}
   \setbox\volboxJPE=\null            \gdef\isvolJPE{F}
   \setbox\preprintboxJPE=\null       \gdef\ispreprintJPE{F}
   \setbox\toappearboxJPE=\null       \gdef\istoappearJPE{F}
   \setbox\inprintboxJPE=\null        \gdef\isinprintJPE{F}
   \setbox\bookboxJPE=\null           \gdef\isbookJPE{F}  \gdef\isinbookJPE{F}
     
   \setbox\bybookboxJPE=\null         \gdef\isbybookJPE{F}
   \setbox\publisherboxJPE=\null      \gdef\ispublisherJPE{F}
}
\def\widestlabel#1{\setbox0=\hbox{#1\enspace}\refindent=\wd0\relax}
\def\ref{\refclearJPE}
\def\no#1{\gdef\isnoJPE{T}\setbox\noboxJPE=\hbox{#1}}
\def\by#1{\gdef\isbyJPE{T}\setbox\byboxJPE=\hbox{#1}}
\def\paper#1{\gdef\ispaperJPE{T}\setbox\paperboxJPE=\hbox{#1}}
\def\yr#1{\gdef\isyrJPE{T}\setbox\yrboxJPE=\hbox{#1}}
\def\jour#1{\gdef\isjourJPE{T}\setbox\jourboxJPE=\hbox{#1}}
\def\pages#1{\gdef\ispagesJPE{T}\setbox\pagesboxJPE=\hbox{#1}}
\def\vol#1{\gdef\isvolJPE{T}\setbox\volboxJPE=\hbox{\bf #1}}
\def\preprint#1{\gdef
\ispreprintJPE{T}\setbox\preprintboxJPE=\hbox{#1}}

\def\book#1{\gdef\isbookJPE{T}\setbox\bookboxJPE=\hbox{\em #1}}
\def\publisher#1{\gdef
\ispublisherJPE{T}\setbox\publisherboxJPE=\hbox{#1}}
\def\inbook#1{\gdef\isinbookJPE{T}\setbox\bookboxJPE=\hbox{\em #1}}
\def\bybook#1{\gdef\isbybookJPE{T}\setbox\bybookboxJPE=\hbox{#1}}
\newdimen\refindent
\refindent=5em
\def\endref{\sfcode`.=1000
 \if T\isnoJPE
\hangindent\refindent\hangafter=1
      \noindent\hbox to\refindent{[\unhbox\noboxJPE\unskip]\hss}\ignorespaces
     \else  \noindent    \fi
 \if T\isbyJPE    \unhbox\byboxJPE\unskip: \fi
 \if T\ispaperJPE \unhbox\paperboxJPE\unskip\period \fi
 \if T\isbookJPE {\it\unhbox\bookboxJPE\unskip}\if T\ispublisherJPE, \else.
\fi\fi
 \if T\isinbookJPE In {\it\unhbox\bookboxJPE\unskip}\if T\isbybookJPE,
\else\period \fi\fi
 \if T\isbybookJPE  (\unhbox\bybookboxJPE\unskip)\period \fi
 \if T\ispublisherJPE \unhbox\publisherboxJPE\unskip \if T\isjourJPE, \else\if
T\isyrJPE \  \else\period \fi\fi\fi
 \if T\istoappearJPE (To appear)\period \fi
 \if T\ispreprintJPE Pre\-print\period \fi
 \if T\isjourJPE    \unhbox\jourboxJPE\unskip\ \fi
 \if T\isvolJPE     \unhbox\volboxJPE\unskip\if T\ispagesJPE, \else\ \fi\fi
 \if T\ispagesJPE   \unhbox\pagesboxJPE\unskip\  \fi
 \if T\isyrJPE      (\unhbox\yrboxJPE\unskip)\period \fi
 \if T\isinprintJPE (in print)\period \fi
\filbreak
}
\normalbaselineskip=12pt
\baselineskip=12pt
\parskip=0pt
\parindent=22.222pt
\beforesectionskipamount=24pt plus0pt minus6pt
\sectionskipamount=7pt plus3pt minus0pt
\overfullrule=0pt
\hfuzz=2pt
\nopagenumbers
\headline={\ifnum\pageno>1 {\hss\tenrm-\ \folio\ -\hss} \else
{\hfill}\fi}
\if F\PSfonts
\font\titlefont=cmbx10 scaled\magstep2

\font\toplinefont=cmr10
\font\pagenumberfont=cmr10
\let\tenpoint=\rm
\else
\font\titlefont=ptmb at 14 pt

\font\toplinefont=cmcsc10
\font\pagenumberfont=ptmb at 10pt
\fi
\newdimen\itemindent\itemindent=1.5em

\def\textindent#1{\indent\llap{#1\enspace}\ignorespaces}
\def\item{\par\noindent
\hangindent\itemindent\hangafter=1\relax
\setitemmark}
\def\setitemindent#1{\setbox0=\hbox{\ignorespaces#1\unskip\enspace}%
\itemindent=\wd0\relax
\message{|\string\setitemindent: Mark width modified to hold
         |`\string#1' plus an \string\enspace\space gap. }%
}
\def\setitemmark#1{\checkitemmark{#1}%
\hbox to\itemindent{\hss#1\enspace}\ignorespaces}
\def\checkitemmark#1{\setbox0=\hbox{\enspace#1}%
\ifdim\wd0>\itemindent
   \message{|\string\item: Your mark `\string#1' is too wide. }%
\fi}
\def\SECTION#1{\vskip0pt plus.2\vsize\penalty-75
    \vskip0pt plus -.2\vsize
    \global\advance\SECTIONcount by 1
    \beforesectionskip\noindent
{\sectionsize\sectiontype \actualnumber.\ #1}
    \EQNcount=1
    \CLAIMcount=1
    \SUBSECTIONcount=1
    \nobreak\sectionskip\noindent}
\catcode`\@=11
\def\@checkfirstline{\ifundefined{nonemptydefs}
\gdef\@haddefs{F}
\immediate\write16{--------Recreating defs.lst}
\else
\gdef\@haddefs{T}
\immediate\write16{--------Using old defs.lst}
\fi
}
\def\@definereally#1#2#3{\expandafter\xdef\csname #1#2\endcsname{#3}\relax}
\def\@writetofile#1#2#3{%
\write\@defs{\string\expandafter\string\xdef\string\csname}\relax
\write\@defs{\string #1#2\string\endcsname{\csname #1#2\endcsname}}}
\gdef\@haddefs{T}
\gdef\@neednewrun{T}
\gdef\@erasedefs{F}
\gdef\@doubly{F}
\newread\@defs
\openin\@defs defs.lst
\ifeof\@defs
\gdef\@haddefs{F}
\immediate\write16{*********Creating defs.lst}
\else
\closein\@defs
\input defs.lst
\@checkfirstline
\fi
\openout\@defs=defs.lst
\write\@defs{\string\def\string\nonemptydefs{}}
\def\@phaseone(#1,#2,#3){\ifundefined{#1#2}
\@definereally{#1}{#2}{#3}
\@writetofile{#1}{#2}{#3}
\else
 \write16{!!!!!doubly defined #1,#2}
 \gdef\@doubly{T}
\fi
}
\def\@phasetwo(#1,#2,#3){\@undefineda{#1#2}
\else
\edef\@firstarg{\csname #1#2\endcsname}
\edef\@secondarg{#3}
\ifnum \@stringcompare{\@firstarg}{\@secondarg} = 0
\else
\immediate\write16{ definition of #1 #2 changed : \@firstarg  --->\@secondarg}
\gdef\@neednewrun{T}
\gdef\@erasedefs{T}
\fi
\fi
\@definereally{#1}{#2}{#3}
\@writetofile{#1}{#2}{#3}}
\def\@undefineda#1{\expandafter\ifx\csname#1\endcsname\relax
{\gdef\@neednewrun{T}
 \immediate\write16{*********** used #1 which was never defined }}}
\if F\@haddefs
 \def\NEWDEF#1#2#3{\@phaseone({#1},{#2},{#3})}
\else
 \def\@neednewrun{F}
 \let\ifundefined=\@undefineda
 \def\NEWDEF#1#2#3{\@phasetwo({#1},{#2},{#3})}
\fi
\def\bye{\if T\@erasedefs
 \write16{**** I suspect doubly defined tokens ****}
 \write16{**** I erase defs.lst now ***********}
 \openout\@defs=defs.lst
\fi
\if T\@doubly
 \write16{**** There are doubly defined tokens *****}
 \write16{**** You have to correct the TeX file and rerun *****}
\fi
\if T\@neednewrun
 \write16{*********NEED ANOTHER RUN************}
\fi
\end}
\def\@stringcompare#1#2{\expandafter\@strcompa#1|===#2|===}
\def\@strcompa#1===#2==={\expandafter\@strcompb#2===#1===}
\def\@strcompb#1===#2==={\@compcont#1\\#2\\}
\def\@compcont#1#2\\#3#4\\{
\csname @comp
 \if #1#3\if #1|same\else contt\fi
   \else diff\fi
  \endcsname #2\\#4\\}
\def\@compcontt#1#2\\#3#4\\{\@strcompb#1#2===#3#4===}
\def\@compsame#1\\#2\\{0}
\def\@compdiff#1\\#2\\{1}
\catcode`\@=\active
\def\tpsi{\tilde\psi}
\def\vcn{\hat v_{\c,n }}
\def\vsn{\hat v_{\s,n }}
\def\hLL{\widehat \LL}
\def\frac#1#2{{#1\over #2}}
\def\geq{\ge}
\let\theta=\vartheta
\let\kappa=\varkappa
\let\epsilon=\varepsilon
\def\Kc{K_{ct}}
\def\hKc{\widehat\Kc}
\def\hK{\widehat K}
\def\HHuu{\tilde\H_{2}^{2,\delta}}
\def\HHq{\tilde\H_{0}^{2,\delta}}
\def\HHbuu{\hat\H_{2}^{2,\delta}}
\def\Huu{\H_{2,\delta}^2}
\def\Zuu{\KK_{1}}
\def\HHvoid{}

\def\d{\kern -0.2em{\rm d}}
\def\dx{\d x\,}
\def\dk{\d k\,}
\def\dm{\d m\,}
\def\dell{\d \ell\,}
\def\L{{\rm L}}
\def\H{{\rm H\kern0.05em}}
\def\tW{\widetilde W}
\def\tc{\hat c}
\def\cc{{c}}
\def\FF{{\cal F}}
\def\AA{{\cal A}}
\def\KK{{\cal K}}
\def\Cbb{{{\cal C}^2_{\rm b,\delta }}}
\def\Cnullb{{{\cal C}^0_{\rm b}}}
\def\tU{\widetilde U}
\def\tu{\tilde u}
\def\tw{\tilde w}
\def\ta{\tilde a}
\def\tr{\tilde r}
\def\WW{{\cal W}_{\beta ,\tc t}}
\def\TT{T}
\def\w{{\rm w}}
\def\r{{\rm r}}
\def\Ustar{U_{\kern-0.15em *\kern+0.15em}}
\def\hPi{\widehat \Pi}
\def\tPi{\widetilde \Pi}
\def\betaA{\beta_{\kern-0.15em A\kern+0.15em}}
\def\hu{\hat u}
\def\ub{\hat u}
\def\hr{\hat r}
\def\SS{{\cal S}}
\def\CC{{\cal C}}
\newfam\funnysy
\font\fffff=cmsy10 at 7pt
\textfont\funnysy=\fffff
\def\bbbbb{\mathchar"0\hexnumber\funnysy0D }
\def\starb{\,*\kern -0.82em{\raise 0.08em\hbox{$\bbbbb$}\,}}
\def\L{{\rm L}}
\def\hS{\widehat S}
\let\truett=\tt
\fontdimen3\tentt=2pt\fontdimen4\tentt=2pt
\def\tt{\hfill\break\null\kern -2truecm\truett *****}
\def\text#1{\leavevmode\hbox{#1}}
\def\UQA{U_{q,a}}
\def\FCQA{F_{c,q,a}}
\def\TC{\tau _{ct}}
\def\MMi{\MM_{\bf i}}
\def\NNi{\NN_{\bf i}}
\let\epsilon=\varepsilon
\let\phi=\varphi
\let\theta=\vartheta
\def\Pc{\hat P_{\rm c}}

\def\Ech{\hat E_{\rm c}^{\rm h}}
\def\Esh{\hat E_{\rm s}^{\rm h}}
\def\Ec{\hat E_{\rm c}}
\def\Es{\hat E_{\rm s}}
\def\ru{\rho^u}
\def\Ru{R^u}
\def\rr{\rho^r}
\def\c{{\rm c}}
\def\s{{\rm s}}
\def\vc{\hat v_{\rm c}}
\def\vs{\hat v_{\rm s}}
\def\uc{\hat u_{\rm c}}
\def\us{\hat u_{\rm s}}
\def\hw{\widehat w}
\def\hv{\hat v}
\let\theta=\vartheta
\let\kappa=\varkappa
\let\epsilon=\varepsilon
\def\ie{{\it i.e.},\ }
\def\hMM{\widehat \MM}

\def\hNN{\widehat \NN}
\def\c{{\rm c}}
\def\s{{\rm s}}
\def\JJ{{\cal T}}
\headline
{\ifnum\pageno>1 {\toplinefont Stability of Modulated Fronts}
\hfill{\pagenumberfont\folio}\fi}
{\titlefont{\centerline{Non-linear Stability of Modulated Fronts}}}
\vskip 0.5truecm
{\titlefont{\centerline{{for the Swift-Hohenberg Equation}}}
\vskip 0.5truecm
{\it{\centerline{J.-P. Eckmann${}^{1,2}$ and G. Schneider${}^3$}}}
\vskip 0.3truecm
{\eightpoint
\centerline{${}^1$D\'ept.~de Physique Th\'eorique, Universit\'e de Gen\`eve,
CH-1211 Gen\`eve 4, Switzerland}
\centerline{${}^2$Section de Math\'ematiques, Universit\'e de Gen\`eve,
CH-1211 Gen\`eve 4, Switzerland}
\centerline{${}^3$ Mathematisches Institut, Universit\"at Bayreuth,
D-95440 Bayreuth, Germany}
}}
\vskip 1cm
{\eightpoint\baselineskip 11pt\narrower
\LIKEREMARK{Abstract}We consider front solutions of the
Swift-Hohenberg equation $\partial _t u\,=\,
-(1+\partial _x^2)^2 u +\epsilon ^2 u -u^3$. These are traveling waves
which leave in their wake a periodic pattern in the laboratory frame.
Using renormalization techniques and a decomposition into Bloch waves,
we show the non-linear stability of these
solutions.
It turns out that this problem is closely related to the
question of stability of the trivial solution for the model problem
$\partial _t u(x,t) = \partial _x^2 u
(x,t)+(1+\tanh(x-ct))u(x,t)+u(x,t)^p$ with
$p>3$. In particular, we show that the instability of the
perturbation ahead of the front is entirely compensated by a diffusive
stabilization which sets in once the perturbation has hit the bulk
behind the front.
}

\LIKEREMARK{Contents}
{\obeylines{
\sec{statement}: Statement of the problem
{\bf Part I. A simplified problem}
\sec{Introduction}: {The model equation}
\sec{FS}: {Function spaces and Fourier transform}
\sec{linear}: {The linear simplified problem}}
\sec{Renormalization}: {The renormalization approach for the simplified problem
\sec{Scaled}: {The scaled linear problem}
\sec{apriori}: {An a priori bound on the non-linear problem}}
\sec{IP0}: {The iteration process}
{\bf Part II. The Swift-Hohenberg equation}
\sec{Bloch waves}: {Bloch waves}
\sec{fulllin}: {The linearized problem}
\sec{unweightsh}: {The unweighted representation}
\sec{weightsh}: {The weighted representation}
\sec{renorm}: {The renormalization process for the full problem}
\sec{Scaledsh}: {The scaled linear problem}
\sec{nonltermsh}: {The scaled non-linear terms}
\sec{integral}: {Bounds on the integrals}
\sec{scaleinisec}: {Bounds on the initial condition}
\sec{apriorish}: {A priori bounds on the non-linear problem}
\sec{IP}: {The iteration process}
}

\SECT{statement}{Statement of the problem}We consider the
Swift-Hohenberg equation
$$
\partial _t u\,=\,
-(1+\partial _x^2)^2 u +\epsilon ^2 u -u^3~,
\EQ{sh}
$$
with $u(x,t)\in\real$, $x\in\real$, $t\ge 0$ and $0<\epsilon  \ll 1$ a
small bifurcation parameter. It has been shown some time ago that a
2-parameter family of (small) spatially periodic solutions exists
which are independent of $t$. These solutions correspond to a periodic
pattern which exists in the laboratory frame. These solutions are of
the form
$$
\UQA(x) \,=\,  A_{q} \cos\bigl((1+\epsilon
q)x+a\bigr)+\OO(\epsilon ^2)~,
$$
which bifurcate from the solution $u\equiv0$.
Here,
$$
A_q\,=\,2\epsilon  (1-4q^2)^{1/2}~.
$$
It is furthermore well-known and proved in [CE90a] that these solutions
are marginally stable for $4|q|^2\le{1\over 3}$, the so-called Eckhaus
stability range ([Eck65]), and that the spectrum of the linearization about
these solutions is all of $\real^{-}$. Finally, after a long time
it was shown in [Schn96] that these solutions are also non-linearly
stable, and this proof was the presented in a slightly different form
in [EWW97].

In another direction, in earlier work of [CE86] and [EW91] traveling wave
solutions of a special kind leaving a {\em fixed} pattern in the
laboratory space were shown to exist, and their linear stability was
studied in [CE87]. Our present paper is concerned with a first proof of
the {\em non-linear} stability of these traveling solutions.

We first describe the traveling solutions. One way to view them is to
write
$$
\UQA(x)\,=\,\HALF\sum_{n\in2\integer+1} A_{q,n}
e^{in((1+\epsilon q)x+a)}
~,
$$
where $A_{q,1}=A_q$ as defined above and $A_{q,-n}=\bar A_{q,n}$,
with $\bar x$ the complex conjugate of $x$.
Here, the $A_{q,n}$ are in fact $\OO(\epsilon ^{|n|})$, and furthermore
$\UQA$ extends to an analytic function.
The modulated front solutions are then of the form
$$
u(x,t)\,=\,\FCQA(x-ct,x)~,
$$
with
$$
\FCQA(\xi,x)\,=\,\HALF\sum_{n\in2\integer+1} W_{c,q,n}(\xi) e^{in((1+\epsilon
q)x+a) }
~.
\EQ{fcqa}
$$
Note that these are {\em not} classical traveling waves of the form
$u(x-ct)$, and
note furthermore
that $\FCQA$ is periodic in its second argument (with period
$2\pi/(1+\epsilon q)$). The modulated front solutions satisfy [CE86,
EW91],
when $c>0$:
$$
\lim_{\xi\to-\infty } W_{c,q,n}(\xi)\,=\,A_{q,n}~,\quad
\lim_{\xi\to\infty } W_{c,q,n}(\xi)\,=\,0~.
$$
These modulated front solutions are constructed with the help
of a center manifold reduction, where all $ W_{c,q,n} $
are determined by  the central modes $W_{c,q,\pm 1}$.
In the reduced four-dimensional system for $W_{c,q,\pm 1}=
W_{c,q,\pm 1}(\xi) $
there is a heteroclinic
connection lying in the intersection of a four-dimensional
stable manifold of the origin and a two-dimensional unstable manifold
of an equilibrium corresponding to $ U_{q,a} $.
Since this is a very robust situation these solutions can be constructed
by some perturbation analysis  from the ones for $ q = 0 $.
For small $\epsilon $ and $ q = 0 $ the  solution $ W_{c,0,1} $ of
the amplitude equation on the center manifold is close
to the real-valued front solution
$ W_{c,0,1}(\xi) = \epsilon B (\epsilon \xi ) = \epsilon B ( \zeta) $
of the equation
$$
4 \partial _\zeta^2 B +  c_B \partial _\zeta B +  B
- 3 B  |B|^2=\,0~,
$$
connecting $ W_{c,0,1} = 0 $ at $\zeta=+\infty $
with $ W_{c,0,1} = A_0 $ at $\zeta=-\infty $.
The constant $c_B$ is given by $ c_B = \epsilon^{-1} c = \OO(1) $.
Our paper deals with the question:
{\em Under which conditions does the solution of \equ{sh}
with initial data $\FCQA(x,x)+v(x)$ converge to $\FCQA(x-ct,x)$ as
$t\to\infty $?}

We will show our results for the case $q=0$ and $a=0$ only, to keep
the notation on a reasonable level.
The extension to arbitrary $a$ is
trivial by translating the origin, while the extension to arbitrary
$q$ satisfying $4|q|^2<{1\over 3}$ necessitates some notational work
and leads to bounds which depend on $q$.
Thus, we will write the periodic solution as
$$
\Ustar (x)\,=\, A \cos x +\OO(\epsilon ^2)~,
\EQ{periodic}
$$
with $A= 2\epsilon $,
and the modulated front (moving with speed $c = \OO(\epsilon) $) as
$$
F_c(\xi,x)\,=\,\HALF \sum_{n\in2\integer+1} W_c(\xi) e^{inx}~.
$$

We describe next the nature of the stability problem.
Consider an initial condition $u_0(x)=F_c(x,x)+v_0(x)$, and let
$u(x,t)$ denote the solution of \equ{sh} with that initial condition.
Since $F_c$ solves \equ{sh}, we find for the evolution of
$v(x,t)\equiv u(x,t)-F_c(x-ct,x)$:
$$
\partial _t v(x,t)\,=\,\bigl(Lv\bigr)(x,t) - 3F_c(x-ct,x)^2 v(x,t)
- 3F_c(x-ct,x) v(x,t)^2
-v(x,t)^3~.
\EQ{form1}
$$
Here, $L=-(1+
\partial _x^2)^2+\epsilon ^2$. We define the translation operator
$\TC$ by $(\TC f)(x)=f(x-ct,x)$, so that \equ{form1} can be written as
$$
\partial _t v\,=\,Lv - 3(\TC F_c)^2 v
- 3(\TC F_c) v^2
-v^3~.
\EQ{form2}
$$
Introduce now $\Kc $ (the difference between the modulated front and
the periodic solution) by
$$
\Kc (x)\,=\,\bigl(\TC F_c\bigr)(x)-\Ustar (x)\,=\,F_c(x-ct,x)-\Ustar (x)~.
\EQ{kcdef}
$$
Note that $\Kc(x)$ vanishes as $x\to-\infty $, and approaches
$U_*(x)$ as $x\to\infty $.
With these notations we can rewrite \equ{form2} as
$$
\eqalign{
\partial _t v \,&=\,  Lv - 3 \Ustar ^2 v - 6 \Ustar  \Kc  v - 3 \Kc ^2 v-3 \Ustar v^2 -v^3
-3 \Kc  v^2\cr
\,&=\,
\MM v+\MMi v +\NN(v)+ \NNi(v)~,
}\EQ{13a}
$$
where
$$
\eqalign{
\MM v \,&=\,Lv -3 \Ustar ^2 v~,\cr
\MMi v \,&=\,-6\Ustar \Kc  v-3 \Kc ^2 v~,\cr
\NN(v) \,&=\,-3\Ustar v^2-v^3~,\cr
\NNi(v) \,&=\,-3\Kc  v^2~.\cr
 }\EQ{13b}
$$
The variables with index ${\bf i}$
vanish with some exponential rate for fixed $ x \in \real $
in the laboratory frame.
They will be seen to be
exponentially ``irrelevant'' in
terms of a renormalization group analysis.
In order to explain this renormalization problem, we will study, in
the next section the model problem
$$
\partial _t u(x,t)\,=\,\partial _x^2 u(x,t)+a(x-ct) u(x,t) + u(x,t)^p
~,
$$
with $a(\xi)=\HALF (1+\tanh\xi)$, and $p>3$.
This problem is nice in its own right.
The similitude will come from the correspondence of
$\MM$ with $\partial _x^2$, and of $\MMi v$ with the term
$a(x-ct)u(x,t)$.
Indeed:
\item{$\bullet$}the first term will be seen to be diffusive in the
laboratory frame,
\item{$\bullet$}the second term will be seen to be irrelevant in the
laboratory frame,
but the first together with the second term will be
exponentially damping in a suitable space of
exponentially decaying functions in a frame moving with a speed close
to $c$.

\noindent
As in previous work [Sa77, BK94, Ga94, EW94]
our analysis will be based on an interplay
of estimates obtained in these two topologies.

Our main results are stated in \clm{main1} for the simplified problem
and in \clm{main2} for the Swift-Hohenberg problem. We not only show
convergence to the front, but give also precise first order estimates
in both cases. As far as possible, the treatment of the two problems
is done in analogous fashion, so that the reader who has followed the
proof of the simplified problem should have no difficulty in reading
the proof for the full, more complicated, problem.
\REMARK An ideal treatment of this problem would necessitate a norm in
a frame moving with the {\em same} speed as the front. Such a space is
needed to study the stability of so-called critical fronts (moving at
the minimal possible speed where they are linearly stable). Achieving
this aim seems to be a necessary step in solving the long-standing
problem of ``front selection'' [DL83], in a case where the maximum
principle [AW78] is not available.
\REMARK The method also applies to more complicated systems,
like hydrodynamic stability problems. A typical example
are the fronts connecting the Taylor vortices with the Couette flow
in the Taylor-Couette problem. These fronts have been constructed in
[HS99]. The stability of the spatially periodic Taylor vortices
has been shown in [Schn98].

\LIKEREMARK{Notation}Throughout this paper many different constants are denoted
with the same symbol $ C $.

\LIKEREMARK{Acknowledgements}Guido Schneider would like to thank for
the kind hospitality at the Physics
Department  of the  University of Geneva. This work is partially
supported by the Fonds National Suisse.
The work of  Guido Schneider is partially supported  by the
Deutsche Forschungsgemeinschaft DFG
under the grant Mi459/2--3.

\SECTIONNONR{Part I. A simplified problem}
\global\advance\SECTIONcount by -1
\vskip -1truecm
\SECT{Introduction}{The model equation}
Let
$$
a(\xi)\,=\,\HALF(1 +\tanh \xi)~.
\EQ{a}
$$
We want to study the equation
$$
\partial _t u(x,t)\,=\,\partial _x^2 u(x,t)+a(x-\cc t)u(x,t)+u(x,t)^p~,
\EQ{uequ}
$$
with $\cc>0$ and $p>3$.
For notational simplicity we assume $ p \in \natural $.

To understand the dynamics of
\equ{uequ} it might be useful to consider the following simplified
problem
$$
\partial_t v(x,t)\,=\,\partial_x^2 v(x,t)+\theta(x-\cc t)v(x,t)~,
\EQ{theta1}
$$
where $\theta(z)=1$ when $z>0$ and $\theta(z)=0$ when $z<0$.
If we go to the moving frame $\xi=x-\cc t$ and let $w(\xi,t)=v(x-\cc t,t)$,
then the equation for $w$ becomes
$$
\partial_t w(\xi,t)\,=\,\partial_\xi^2 w(\xi,t) +\cc\partial_\xi
w(\xi,t)+ \theta(\xi)w(\xi,t)~.
\EQ{theta2}
$$
For $x>0$, we have $\theta(x)=1$ and hence the corresponding
characteristic polynomial for \equ{theta2} (in momentum space) is
$$
-k^2 +i\cc k +1~,
$$
while for $x<0$, we have $\theta(x)=0$ with its corresponding
polynomial
$$
-k^2+i\cc k~.
$$
Thus, we expect the solution to be exponentially unstable ahead of the
front, \ie for $x>0$, and diffusively stable behind the front.
If we consider an initial condition $v_0(\xi)$ localized near
$\xi=\xi_0>0$,
and of amplitude $A$, then we expect the amplitude to grow like
$e^{t}A$ until $t=t_*=\xi_0/\cc$, when this perturbation ``hits'' the back
of the front (in the moving frame), or, in other words, when the back
of the front hits the perturbation (in the laboratory frame).
Thus, the perturbation does not grow larger than $Ae^{\xi_0 /\cc}$.
We use this in the following way. Assume that the amplitude at $\xi>0$
is bounded
by $Ae^{-\beta \xi}$. Then, ignoring diffusion, we find that the
contribution to the amplitude at the origin at time $t=\xi_0/\cc$ is
bounded by
$$
\int _0^{\xi_0} \d \xi\, Ae^{\xi{(1-\beta \cc)/ \cc}}~.
$$
Clearly, if $\beta \cc >1$, the initial perturbations are sufficiently
small for the total effect at the origin (in the moving frame) to be
small.

Once
this has happened, a second epoch starts where the perturbation is
behind the front. Then, due to the diffusive behavior, the amplitude
will go down as
$$
{C \over (t-t_*+1)^{1/2}}~.
$$
These considerations will be used in the choice of topology below.
\SUBSECT{FS}{Function spaces and Fourier transform}We start the
precise analysis
and will work in Fourier space and revert to the $x$-variables only at
the end of the discussion.
We define the Fourier transform by
$$
\bigl(\FF f\bigr)(k)\,=\,{1\over 2\pi} \int \dx f(x) e^{-ikx}~.
$$
\LIKEREMARK{Notation}If $f$ denotes a function, then $\tilde f$ is
defined by $\tilde f=\FF f$, and if $\AA$ is an operator, then $\tilde
\AA$ is defined by $\tilde \AA= \FF \AA \FF^{-1}$.
We also use the notation $\tilde f* \tilde g$ for the convolution
product
$\widetilde{fg}=(\tilde f *\tilde g)(k)=\int \d\ell \tilde
f(k-\ell)\tilde g(\ell)$. Finally,
$\widetilde\TT_\zeta$ denotes the conjugate of translation:
$$
(\widetilde\TT_\zeta \tilde f)(k)\,=\, e^{-i\zeta k} \tilde f(k)~,
\EQ{tau}
$$
so that the Fourier transform of $\bigl(\TT_\zeta f\bigr)(x)=f(x-\zeta)$ is
$$
\FF\TT_\zeta f\,=\,\widetilde\TT_\zeta \FF f~.
$$
The relation ([Ta97])
$$
k^{\alpha} \partial_k^{\beta} (\FF f)(k) = (-1)^{\beta}
\FF (\partial^{\alpha}_x x^{\beta} f)(k )
$$
motivates the introduction of the
following norms: We fix a small $\delta >0$ and
define
$$
\|\tilde f\|_{\HHuu}\,=\,\left (\sum_{j,\ell=0}^2\delta ^{2(\ell+j)}\int \dk
|\partial_k^j (k^{\ell}\tilde f(k))|^2  \right )^{1/ 2} ~.
\EQ{hhuunorm}
$$
The dual norm to this is
$$
\| f\|_{\Huu}\,=\,\left (\sum_{j,\ell=0}^2\delta ^{2(\ell+j)}\int \dx
|\partial_x^\ell  f(x)|^2 x^{2j} \right )^{1/ 2} ~.
\EQ{huunorm}
$$
Parseval's inequality immediately leads to:
$$
\| f\|_{\Huu}\,=\,\|\FF f\|_{\HHuu}~,
$$
and, for some constant $C$ independent of $1\ge\delta>0 $,
$$
\eqalign{
\| fg\|_{\Huu}\,&\le\,C\| f\|_{\Huu}\| g\|_{\Huu}~,\cr
\| \tilde f*\tilde g\|_{\HHuu}\,&\le\,C\| \tilde f\|_{\HHuu}\|\tilde
g\|_{\HHuu}~.\cr
}
\EQ{fg}
$$
Finally, we shall also need the inequality
$$
\|\tilde f * \tilde g\|_{\HHuu} \,\le\, \|f\|_{\Cbb} \|\tilde g\|_{\HHuu}~,
\EQ{fg2}
$$
where
$$
\|f\|_{\Cbb}=\sum_{j=0}^2 \delta ^j \sup_{x\in\real}|\partial_x^j
f(x)|~.
\EQ{cbbdef}
$$
This follows from
$$
\|\tilde f * \tilde g\|_{\HHuu} \,=\,\|f \cdot g\|_{\Huu}
\,\le\,\|f\|_{\Cbb}\| g\|_{\Huu} \,=\,\|f\|_{\Cbb} \|\tilde
g\|_{\HHuu}~,
$$
where the inequality above is a direct consequence of the definition
of $\HHuu$.
\LIKEREMARK{Notation}In the sequel, we will always write
$\|\cdot\|$ instead of $\|\cdot\|_{\HHuu}$. Thus this is our default norm.

We define the map $\WW$ by
$$
(\WW  f)(\xi)\,=\,f(\xi+\tc t)e^{\beta \xi}~,
\EQ{WWdef}
$$
where $\beta \in(0,\beta _*)$ and $\tc\in (0,\cc)$ will be fixed later.
The Fourier conjugate of this operator then satisfies
$$
\bigl(\widetilde \WW  \tilde f\bigr)(k )\,\equiv\,\bigl(\FF \WW\FF^{-1} \tilde f\bigr)(k
)\,=\,e^{i(k+i\beta )\tc t}\tilde f(k+i\beta  )~,
\EQ{WWhatdef}
$$
as one sees from the following equalities:
$$
\eqalign{
2\pi(\widetilde \WW  \tilde f)(k )\,&=\,\int \d \xi\, e^{-ik\xi }\bigl(\WW
f\bigr)(\xi)\cr
\,&=\,\int \d \xi\, e^{-ik \xi} f(\xi+\tc t)e^{\beta \xi}\cr
\,&=\,\int \d \xi\, e^{-i(k+i\beta )\xi } f(\xi+\tc t)\cr
\,&=\,\int \d \xi\, e^{-i(k+i\beta )(\xi-\tc t)} f(\xi)\cr
\,&=\,2\pi e^{i(k+i\beta )\tc t}\tilde f(k+i\beta  )~.\cr
}
$$
This calculation also shows that {\em if $f(\xi)e^{\beta_* \xi}\in\Huu$
for $ f \in \Cbb $,
then $\widetilde \WW \tilde f$
extends to
an analytic function in $\{0>\Im k > -\beta_*\}$ and $(\widetilde\WW
\tilde f)(\cdot
-i\beta )\in\HHuu$ for all $\beta  \in[0,\beta_* )$.}
\REMARK Since the norms for different $ \delta $
are equivalent, all theorems throughout this paper
can also be formulated in a version with $ \delta = 1 $.

\SECT{linear}{The linear simplified problem}In this section we study the
linearization of equation \equ{uequ}:
$$
\partial_t U(x,t)\,=\,\partial_x^2 U(x,t)+a(x-\cc t) U(x,t)~.
\EQ{ulin}
$$
The function $a$ is given as
$$
a(\xi)\,=\,\HALF(1 +\tanh \xi)~,
\EQ{a3}
$$
but our methods will work for many other functions.
The crucial property we need is the existence of a $\beta _*>0$
such that $a(\xi)e^{-\beta \xi}$ satisfies
$$
\|\xi\mapsto a(\xi)e^{-\beta \xi}\|_{\Huu}\,\le\,C~,
\EQ{a4}
$$
for all $\beta
\in(0,\beta _*)$. For the case of \equ{a3} we can take $\beta _*=2$.
The Fourier transform $\ta$ of $a$ is therefore a tempered
distribution which is the boundary value of a function
(again called $\ta$) which is analytic in the strip $\{z~|~ 0> \Im z >
\beta _*\}$.
Furthermore, there is a $K$ such that, for all $\delta \in (0,1] $,
$$
\|a\|_{\Cbb}\,\le\,1+K\delta ~,
\EQ{adelta}
$$
since
$$
\sup_{x\in \real} |a(x)| \,\le\,1~.
\EQ{a5}
$$
The bound \equ{a5} will be tacitly used later.

The next proposition describes how solutions of \equ{ulin} tend to 0
as $t\to\infty $. We write $U_t(x)$ for $U(x,t)$ and use similar
notation for other functions of space and time.
\CLAIM{Proposition}{lin}Assume that there are a $\beta$  and a $\tc\in(0,c)$
such that
$\beta^2- \beta \tc +1\equiv-
2\gamma <0$. Then there exists a $\delta\in(0,1]$
such that the following holds. Assume that $U_0\in \Huu$ and that
$W_0(\xi)=\bigl({\cal W}_{\beta ,0}U_0\bigr )(\xi)=U_0(\xi)e^{\beta
\xi}\in \Huu$. {\rm (These conditions are independent of
$\delta>0$.)}
Then the solution $U_t(x)=U(x,t)$ of \equ{ulin} with initial data
$U_0$ exists for all $t>0$ and with $\tilde\psi(k)=e^{-k^2}$
the rescaled solution $ \tilde V(k,t) = \tilde U(k t^{-1/2},t) $
satisfies
$$
\|\tilde V_t- \tilde U_0(0)
\tilde\psi\|_{\HHuu} \,\le\, {C\over (1+t)^{1/2}} \|\tilde U_0\|_{\HHuu}~.
\EQ{res1}
$$
The function $\tW _t= \widetilde \WW \tilde U_t$ satisfies
$$
\|\tW _t\| _{\HHuu}\,\le\,C e^{-3\gamma t/2}\|\tW _0\|_{\HHuu}~.
\EQ{res2}
$$
The constant $C$ does not depend on $U_0$.

\REMARK Note that it is optimal to choose
$\tc$ arbitrarily close to $c$.

\PROOF First of all, we rewrite the equation \equ{ulin} for $U_t$ in terms of
$\tU_t$ and $\tW _t$:
The equation for $W_t=\WW U_t$ is
$$
\eqalign{
\partial _t W(\xi,t)\,&=\,
\partial _\xi^2 W(\xi,t) +(\tc-2\beta )\partial _\xi
W(\xi,t)\cr&+a(\xi-(c-\tc)t)W(\xi,t) + (  \beta ^2-\beta \tc)
W(\xi,t)~.\cr
}
\EQ{wequ0}
$$
Taking Fourier transforms, we then find, omitting the argument $k$ and
using the notation of \equ{tau}:
$$
\eqalignno{
\partial_t \tU_t \,&=\, -k^2\tU_t+ (\widetilde\TT_{c t
}\tilde a)*\tU_t  ~,\NR{hatulin}
\partial_t \tW_t \,&=\,\bigl(\beta ^2-\beta \tc -k^2 +ik(\tc-2\beta
)\bigr)\tW_t +(\widetilde\TT_{(c-\tc)t}\tilde a)* \tW_t ~.\NR{hatwlin}
}
$$
It is at this point that the simultaneous choice of two
representations for the solution and their associated topologies is
crucial.

We first show that $\tW _t$ converges to 0, \ie we show
\equ{res2}.
We find from \equ{fg2}:
$$
\|(\widetilde\TT_\zeta \tilde a)*\tilde f\|_{\HHvoid}
\,\le\,\|a(\cdot -\zeta)\|_{\Cbb}\cdot \|\tilde f\|_{\HHvoid}
\,=\,\|a\|_{\Cbb}\cdot \|\tilde f\|_{\HHvoid}~.
\EQ{af}
$$
Therefore, \equ{adelta} implies
$$
\|(\widetilde\TT_{(c-\tc)t}\tilde a)*\tW _t\|_{\HHvoid}  \,\le\,
(1+K\delta)\|\tW _t\|_{\HHvoid}~,
$$
and we get from \equ{hatwlin} the bound
$$
\HALF\partial _t \|\widetilde W_t\|_{\HHvoid}^2\,\le\,(\beta
^2-\beta \tc +1+K\delta+K_1 \delta ) \|\widetilde W_t\|^2_{\HHvoid}
~,
$$
for a constant $ K_1 $ independent of $ \delta \in (0,1] $.
The term $ K_1 \delta $ comes from the derivatives in the norm 
$ \| \cdot \|_{\HHuu} $. We choose $ \delta > 0 $ so small
that 
$$
\beta^2 - \beta \tc + 1 + (K+K_1) \delta \leq -3\gamma/2 ~.
$$
Integrating over $t$ we get from the choice of $\beta $, $\delta$, and $\tc$:
$$
\|\tW _t\|_{\HHvoid}\,\le\,e^{-3\gamma t/2}\|\tW _0\|_{\HHvoid}~.
\EQ{Wbound}
$$
Thus, we have shown Eq.\equ{res2}.

Next, we study $\tU$.
From \equ{WWhatdef} and deforming the contour of integration, we get
$$
\eqalign{
\bigl(\bigl(\widetilde\TT_{\zeta}\tilde a\bigr)*\tilde f\bigr)(k)
\,&=\, \int \d \ell\, e^{-i\zeta (k-\ell)}\tilde
a(k-\ell)\tilde f(\ell)\cr
\,&=\, \int \d \ell\, e^{-i\zeta (k-\ell)}\tilde
a(k-\ell){\bigl(\widetilde\WW \tilde f\bigr)}(\ell-i\beta)e^{-i\ell\tc t}
\cr
\,&=\, \int \d \ell\, e^{-i\zeta (k-\ell-i\beta )}\tilde
a(k-\ell-i\beta ){\bigl(\widetilde\WW\tilde  f\bigr)}(\ell)e^{-i(\ell+i\beta )\tc t}
\cr
\,&=\, e^{-\beta (\zeta -\tc t)}\int \d \ell\, e^{-i\zeta (k-\ell)}\tilde
a(k-\ell-i\beta ){\bigl(\widetilde\WW\tilde  f\bigr)}(\ell)e^{-i\ell \tc t}
~.\cr
}
\EQ{taa}
$$
Let $\tilde h(k)=e^{-ictk}\tilde a(k-i\beta )$ and $\tilde g(k)=
e^{-ik\tc t}\bigl(\widetilde\WW \tU_t\bigr)(k)=e^{-ik\tc t}\tW_t(k)$. Then
\equ{taa} implies
$$
\bigl(\widetilde\TT_{ct}\ta\bigr)*\tU_t\,=\, e^{-\beta (c-\tc )t} \tilde h* \tilde g~.
$$
{}From this we conclude that
$$
\eqalign{
\|\bigl(\widetilde\TT_{c t}\tilde a\bigr)*\tU _t\|_{\HHvoid}
\,&=\,
e^{-\beta (c -\tc) t}\| \tilde h* \tilde g\|_{\HHvoid}\cr
\,&\le\,e^{-\beta (c-\tc)t}\| \tilde h\|_{\HHvoid} \,\,\|\tilde g\|_{\HHvoid}
\cr
\,&\le\,C (1+t c)^2(1+t \tc)^2e^{-\beta (c-\tc)t}\|\tW _t\|_{\HHvoid}~.
}\EQ{tab}
$$
On the other hand, from \equ{res2} we know that $\|\tW_t\|_{\HHvoid}$
stays bounded (it actually decays exponentially), and thus the
evolution equation for
$\tU_t $ is of the form
$$
\partial _t \tU _t(k)\,=\,-k^2 \tU _t (k) + \tilde h(k,t)(1+t
c)^2(1+t\tc)^2 e^{-\beta (c-\tc)t}~,
$$
with $\|\tilde h(\cdot,t)\|_{\HHvoid}$ uniformly bounded in $t$.
Since, by construction, $\tc<c$,
we conclude that \equ{res1} holds, using well-known arguments
which will be made explicit
in the proof of \clm{main1}.
The proof of
\clm{lin} is complete.\QED
\SECT{Renormalization}{The renormalization approach for the simplified
problem}We consider now
the non-linear problem \equ{uequ} and its related version for
$\tw_t=\widetilde \WW \tilde u_t= \FF \WW u_t$
in Fourier space. It takes the form
$$
\eqalign{
\partial_t \tu_t\,&=\, -k^2 \tu_t
+ \bigl(\widetilde\TT_{c t} \ta) * \tu_t + \tu_t^{*p}~,\cr
\partial_t \tw_t \,&=\, \bigl(\beta ^2-\beta \tc -k^2 +ik(\tc-2\beta
)\bigr)\tw_t+\bigl(\widetilde\TT_{(c-\tc)t}\tilde a)*  \tw_t + \tu_t^{*(p-1)} * \tw_t~.\cr
}\EQ{bothnon}
$$

Let $M_\beta $ be the operator of multiplication: $(M_\beta
f)(x)=e^{\beta x}f(x)$.
Choose the constants $\hat c$,  and $\beta $  such that they
satisfy
as before
$$
0\,>\,-2 \gamma\,=\,\beta ^2- \beta\tc +1~,
$$
and fix them henceforth.
Our main result for the simplified problem is:
\CLAIM{Theorem}{main1}There are positive
constants $R$, $C$ and $ \delta \in (0,1] $ such that the following
holds:
Assume $\|u_0\|_{\Huu}+\|M
_\beta u_0\|_{\Huu} \le R$. Then the solution $u_t$ of \equ{uequ} with
initial condition $u_0$ converges to a Gaussian in the sense that
there is a constant $A_*=A_*(u_0)$ such that with $\tilde\psi(k)=e^{-k^2}$
the rescaled solution $ \tilde{v}(k,t) = \tu ( k t^{-1/2} ,t) $
satisfies
$$
\|\tilde{v}_t-A_* \tilde\psi\|_{\HHuu}\,\le\,{CR\over (t+1)^{1/2}}~.
\EQ{mainineq}
$$
Furthermore,
$$
\|\tw_t\|_{\HHuu}\,=\,\|\FF \WW u_t\|_{\HHuu}\,\le\,CR
e^{-\gamma t}~.
$$

We shall use the renormalization technique of [BK92]
to show that $\tu_t$ and
$\tw_t$ behave (as $t\to\infty $) essentially in the same
way as their linear counterparts $\tU_t$ and $\tW_t $ from the previous
section. This technique consists, see [CEE92], in pushing forward the
solution for some time and then rescaling it. This process makes the
effective non-linearity smaller at each step, so that in the end the
convergence properties of the linearized problem are obtained.

We fix $0<\sigma\le1$ and introduce:
$$
\eqalign{
\bigl(\tilde \LL \tilde f\bigr)( \kappa)\,&=\, \tilde f( \sigma \kappa)~.
}
\EQ{mndef}
$$
This is again a linear change of coordinates in function space.
Note that
$$
\eqalign{
\bigl(\tilde \LL (\tilde f*\tilde g)\bigr)(\kappa)\,&=\,
\int \d\kappa'\, \tilde f(\sigma \kappa-\kappa') \tilde g(\kappa')\cr
\,&=\,\sigma\int \d(\sigma^{-1}\kappa')\, \tilde f(\sigma
\kappa-\sigma\sigma^{-1}\kappa') \tilde g(\sigma\sigma^{-1}\kappa')\cr
\,&=\,\sigma \bigl((\tilde \LL \tilde f)*(\tilde \LL \tilde g)\bigr)(\kappa)~.\cr
}\EQ{id1}
$$
Furthermore,
$$
\bigl(\tilde \LL  (\widetilde\TT_\zeta \ta)\bigr) (\kappa)\,=\,
e^{i\zeta\sigma\kappa} \ta(\sigma\kappa)
\,=\,\bigl(\widetilde\TT_{\sigma  \zeta} (\tilde \LL \ta) \bigr)(\kappa)~,
$$
and therefore we have
$$
\tilde \LL  \bigl((\widetilde\TT_\zeta \ta) * \tilde
f\bigr)\,=\,\sigma
( \widetilde\TT_{\sigma
\zeta} \tilde \LL \ta )*(\tilde \LL  \tilde f)~.
\EQ{id2}
$$

We next define
$$
\eqalign{
\tu_{n,\tau }(\kappa)\,&=\,\bigl(\tilde\LL^n \tu\bigr)(\kappa,\sigma^{-2n}\tau)\,=\,\tu(\sigma^n\kappa,\sigma^{-2n}\tau)~,\cr
\tw_{n,\tau }(\kappa)\,&=\,e^{-\gamma \sigma^{-2n} \tau}
\bigl(\tilde\LL^n \tw\bigr)(\kappa,\sigma^{-2n}\tau)\,=\,
e^{-\gamma \sigma^{-2n} \tau} \tw(\sigma^n\kappa,\sigma^{-2n}\tau)~,\cr
}
$$
so that this corresponds to an additional rescaling of the time axis.
Note that
$$ \tw_{n,\sigma^2 }(\kappa) = e^{-\gamma \sigma^{-2n} \sigma^2}
\tw(\sigma^n\kappa,\sigma^{-2n}\sigma^2 ) = \tw_{n-1,1 }(\sigma^{-n}
\kappa)~.
$$
We also let $\ta_n=\tilde\LL^n\ta$.
From \equ{id1}, \equ{id2}, and $\partial_\tau =
\sigma^{-2n}\partial_t$
we find easily that \equ{bothnon}
transforms to the system (omitting the argument $\kappa$):
$$
\eqalignno{
\partial_\tau \tu_{n,\tau}\,&=\, -\kappa^2 \tu_{n,\tau}
+ \sigma^{-n} (\widetilde\TT_{c \sigma^{-n}\tau } \ta_n) *
\tu_{n,\tau} + \sigma^{n(p-3)}\tu_{n,\tau}^{*p}~,\NR{bothnonn1}
\partial_\tau \tw_{n,\tau} \,&=\, \bigl((\beta ^2-\beta \tc+\gamma)\sigma^{-2n}
-\kappa^2
+i\kappa(\tc-2\beta
)\sigma^{-n}\bigr)\tw_{n,\tau}\NR{bothnonn2}
&~~+\sigma^{-n}(\widetilde\TT_{(c-\tc)\sigma^{-n}\tau}\ta_n)*  \tw_{n,\tau} +
\sigma^{n(p-3)}\tu_{n,\tau}^{*(p-1)} * \tw_{n,\tau}~.\cr
}
$$
We see that under these
rescalings the coefficients of the non-linear terms go to 0 as
$n\to\infty $. We will now
put this observation into more mathematical form.

The equation \equ{bothnon} is of the form $\partial _t X_t=
L\bigl(X_t\bigr)+\NN\bigl(X_t\bigr)$, where $L$ contains the linear
parts with the exception of those depending on $\ta_n$ and $\NN$ denotes
the other terms. We can write the solution as
$$
X_t\,=\,e^{(t-t_0)L}X_{t_0}+\int_{t_0}^t \d s\,e^{(t-s) L} \NN(X_s)~.
$$
Going to the rescaled variables $X_{n,\tau}$, and taking
$t_0=\sigma^{-2(n-1)} $
and $t=\sigma^{-2n} \tau$, we can express this (for the $\tu$) as follows.
The equation \equ{bothnonn1} leads to
$$
\eqalign{
\tu_{n,\tau}(\kappa)\,&=\,e^{-\kappa^2(\tau-\sigma^2)}\tu_{n,\sigma^2}(\kappa)\cr &~~+
\int_{\sigma^2 }^{\tau}\d\tau'\,e^{-\kappa^2(\tau-\tau')}\bigg(
\sigma^{-n} (\widetilde\TT_{c \sigma^{-n}\tau' } \ta_n) *
\tu_{n,\tau'} + \sigma^{n(p-3)}\tu_{n,\tau'}^{*p}\bigg)(\kappa)~.\cr
}
\EQ{new1}
$$
Similarly, we rewrite \equ{bothnonn2} as
$$
\eqalign{
\partial_\tau \tw_{n,\tau} \,&=\,
\tilde G_{n,\tau}\tw_{n,\tau} +
\sigma^{n(p-3)}\bigl(\tu_{n,\tau}^{*(p-1)} * \tw_{n,\tau}\bigr)~,\cr
}
$$
where $\tilde G_{n,\tau }$ is defined, {\it cf.} \equ{bothnonn2}, by
$$
\bigl(\tilde G_{n,\tau }\tilde f\bigr)(\kappa)=\bigl((\beta ^2-\beta \tc+\gamma)\sigma^{-2n}
-\kappa^2
+i\kappa(\tc-2\beta
)\sigma^{-n}\bigr)\tilde f(\kappa)+\sigma^{-n}\bigl
((\widetilde\TT_{(c-\tc)\sigma^{-n}\tau}\ta_n)*  \tilde f \bigr )(\kappa)~.
$$
The solution of
the linear evolution equation $\partial _\tau
\tilde f_{n,\tau}=\tilde G_{n,\tau}\tilde f_{n,\tau}$
is nothing but \equ{hatwlin} in a new coordinate
system.
We write the solution as $\tilde f_{n,\tau}=\tilde
S_{n,\tau,\tau'}\tilde f_{n,\tau'}$.
Then, in analogy to \equ{new1} we get
$$
\tw_{n,\tau
}(\kappa)\,=\,\bigl(\tilde S_{n,\tau,\sigma^2}\tw_{n,\sigma^2}\bigr)(\kappa)
+\sigma^{n(p-3)}\int_{\sigma^2}^\tau  \d\tau '\,
\biggl(\tilde S_{n,\tau,\tau'}\bigl(\tu_{n,\tau'}^{*(p-1)} *
\tw_{n,\tau'}\bigr)\biggr) (\kappa)~.
\EQ{new3}
$$
\REMARK The proof of \clm{main1} is divided into several steps:
In \clm{Snn} below,
we improve first the inequalities
for the exponentially damped part in scaled variables.
Then in \clm{apriori} a priori estimates for the solutions
of \equ{new1} and \equ{new3} are established.
With these a priori bounds we show \clm{decay}. From these results,
\clm{main1} will follow rather simply by a contraction argument.
\SUBSECT{Scaled}{The scaled linear problem}Here, we derive the
essential bounds on the influence of the term
$a(x-ct)u(x,t)$ in the equation for $ \tw_t $
under the scalings introduced above.
Note first that, from definition \equ{hhuunorm} and \equ{mndef}, we have
$$
\|\tilde\LL \tilde f\|_{\HHvoid}^2\,=\,
\sigma^{-1}\int \d (\sigma\kappa)\,
\sum_{j,\ell=0}^2 \delta^{2\ell}\sigma^{-2\ell} \sigma^{2j}
|(\partial^j \tilde f)(\sigma\kappa)|^2(\sigma\kappa)^{2\ell} ~.
$$
From this we conclude immediately that for $0<\sigma<1$:
$$
\|\tilde\LL \tilde f\|_{\HHvoid}\,\le\, \sigma^{-5/2} \|\tilde f\|_{\HHvoid}~~~\text{and}~~~
\|\tilde\LL^{-1} \tilde f\|_{\HHvoid}\,\le\, \sigma^{-3/2} \|\tilde f\|_{\HHvoid}~.
\EQ{Lnbound}
$$
We next bound
$\tilde S_{n,\tau,\tau '}$.
Recall that
we are assuming
$\beta ^2-\beta \tc+1 = -2 \gamma<0$.
\CLAIM{Lemma}{Snn}For all $ \epsilon'\in (0,1) $ there exists a $ C_{\epsilon'}
> 0 $
such that for $1>\tau >\tau '\ge0$ one has
$$
\eqalign{
\|\tilde S_{n,\tau,\tau'}\tilde  f\|_{\HHvoid}
\,&\le\,  C_{\epsilon'} \sigma^{-\epsilon' n}e^{-\gamma
\sigma^{-2n}(\tau -\tau ')/ 2}\|\tilde f\|_{\HHvoid}~,\cr
}
\EQ{sbound}
$$
for all $ n \in\natural  $.

\PROOF We consider the equation $\partial_\tau \tilde f_{\tau } =
\tilde G_{n,\tau
} \tilde f_{\tau }$, whose solution is $\tilde f_{\tau }=\tilde S_{n,\tau,\tau'} \tilde f_{\tau' }$:
$$
\partial_\tau  \tilde f_{\tau} \,=\, \tilde \lambda_n \tilde f_{\tau}
+\sigma^{-n}(\widetilde\TT_{(c-\tc)\sigma^{-n}\tau}\ta_n)* \tilde   f_{\tau}~,
\EQ{bothnonn2a}
$$
where $\tilde \lambda_n$ is the operator of multiplication by
$$
\tilde \lambda_n(\kappa) \,=\, (\beta ^2-\beta \tc+\gamma)\sigma^{-2n}
-\kappa^2
+i\kappa(\tc-2\beta
)\sigma^{-n}~.
$$
The variation of constant formula yields
$$
\tilde  f_{\tau} = e^{\lambda_n (\tau - \tau')}\tilde   f_{\tau'}
+ \int^{\tau}_ {\tau'}\d s\, e^{\lambda_n (\tau - s)}
\sigma^{-n}(\widetilde\TT_{(c-\tc)\sigma^{-n}s}\ta_n)*\tilde    f_{s}~.
$$
We now introduce the norm
$$
\|\tilde f\|_{\HHq}^2\,=\,\sum_{j=0}^2\delta ^{2j}\int \dk
|\partial_k^j \tilde f(k)|^2  ~,
$$
and its dual
$$
\|f\|_{{\rm H}_{2,\delta}^0}^2\,=\,\sum_{j=0}^2\delta ^{2j}\int \dx
|x^j  f(x)|^2  ~.
$$
We use
$$
\| e^{\lambda_n \tau} \tilde f \|_{\HHq}
\leq \| e^{\lambda_n \tau}  \|_{\Cbb}
\| \tilde f \|_{\HHq}~,
$$
and
$$
\eqalign{
\| e^{\lambda_n \tau}  \|_{\Cbb} & \leq
\| e^{\lambda_n \tau}  \|_{\Cnullb} +
\delta \| \lambda_n' \tau e^{\lambda_n \tau}  \|_{\Cnullb}
+ \delta^2 \| \lambda_n'' \tau e^{\lambda_n \tau}  \|_{\Cnullb}
+ \delta^2 \| (\lambda_n' \tau)^2 e^{\lambda_n \tau}  \|_{\Cnullb}
\cr
& \leq C_{\chi,\delta}
e^{(\beta^2 - \beta \tc + \gamma + \chi) \sigma^{-2n} \tau}~,
}
$$
for every $ \chi > 0 $, where the $ C_{\chi,\delta} $
are constants independent of $ \sigma $ depending only
on  $ \chi $ and $ \delta $. They have the property that
$ \lim_{\delta \rightarrow 0} C_{\chi,\delta} = 1 $ for fixed $ \chi  $.
We choose $ \chi = \gamma/4 $
and  find
$$
\eqalign{
\| \tilde f_{\tau}\|_{\HHq }& \,\leq\,
C_{\gamma/4,\delta} e^{(\beta ^2-\beta \tc+5\gamma/4)
\sigma^{-2n} (\tau - \tau')}\| \tilde
f_{\tau'} \|_{\HHq } \cr &
+ C_{\gamma/4,\delta}
\int^{\tau}_ {\tau'}\d s\, e^{(\beta ^2-\beta \tc+5\gamma/4)
\sigma^{-2n} (\tau - s)}
 \sigma^{-2n} \|a_n \|_{\Cnullb}\,  \| \tilde  f_{s} \|_{\HHq }~,
}
$$
since
$$
\eqalign{
\|(\tilde T_\zeta \FF
a_n)*\tilde f\|_{\HHq} & \,=\,
\sigma^{-n}
\|a_n(\cdot-\zeta)\FF^{-1}\tilde
f\|_{{\rm H}_{2,\delta}^0} \cr & \,\le\,
\sigma^{-n}
\|a_n(\cdot-\zeta)\|_{\Cnullb}\,\|\FF^{-1}
\tilde f\|_{{\rm H}_{2,\delta}^0}\,=\,
\sigma^{-n}
\|a_n\|_{\Cnullb}\,\|
\tilde f\|_{\HHq}~.}
$$
Using $ \|a_n \|_{\Cnullb} = 1 $ and applying Gronwall's inequality
to
$  e^{-(\beta ^2-\beta \tc+5\gamma/4)
\sigma^{-2n} \tau} \| \tilde f_{\tau}\|_{\HHq } $
we get
$$
 e^{-(\beta ^2-\beta \tc+5\gamma/4)
\sigma^{-2n} \tau} \| \tilde f_{\tau}\|_{\HHq }
\,\leq\,
C_{\gamma/4,\delta} e^{C_{\gamma/4,\delta}\sigma^{-2n}
(\tau-\tau')}\| \tilde f_{\tau'}\|_{\HHq }~,
$$
or equivalently,
$$
\|\tilde  f_{\tau}\|_{\HHq }
\,\leq\,
C_{\gamma/4,\delta} \|\tilde  f_{\tau'}\|_{\HHq }
e^{(\beta ^2-\beta \tc+5\gamma/4+ C_{\gamma/4,\delta})
\sigma^{-2n}(\tau-\tau')}~.
\EQ{first}
$$
We choose $ \delta \in (0,1] $ so small that
$ C_{\gamma/4,\delta} < \gamma/4+1 $.
This proves the assertion of \clm{Snn} for the $\HHq$ norm.

We next use the regularizing character of $-\kappa^2$ to
prove the bound in $\HHuu$.
Let $ \tilde q(\kappa) = \kappa $. Then
$$
\eqalign{
\|\tilde  q  \tilde f_{\tau}\|_{\HHq }  \,&\,\leq\, \,
C e^{(\beta ^2-\beta \tc+3 \gamma/2)\sigma^{-2n} (\tau - \tau')}
\|  \tilde q  \tilde f_{\tau'} \|_{\HHq } \cr
& ~+ C \int^{\tau}_ {\tau'}\d s\, e^{(\beta ^2-\beta \tc+3\gamma/2)
\sigma^{-2n} (\tau - s)}
 \sup_{\kappa \in \real} | e^{-\kappa^2 (\tau -s)} \kappa |
\,\,
  \sigma^{-2n} \, \|a_n \|_{\Cnullb}
  \|  \tilde f_{s} \|_{\HHq }~.\cr
}
$$
Using the estimate \equ{first} for $ \|  f_{s} \|_{\HHq } $
we get
$$
\| \tilde q  \tilde f_{\tau}\|_{\HHq }
\,\leq\,
C \| (1+|\tilde q|)  \tilde f_{\tau'}\|_{\HHq }
e^{(\beta ^2-\beta \tc+1+3\gamma/2) \sigma^{-2n} (\tau-\tau')}
\max\bigl (1,(\tau-\tau')^{1/2}\bigr )~.
\EQ{second}
$$
To bound the second power of $\tilde q$, choose $ \epsilon' \in (0,1) $. Then
$$
\eqalign{
\| \tilde q^2  \tilde f_{\tau}\|_{\HHq } \,&\,\leq\,\,
e^{(\beta ^2-\beta \tc+3\gamma/2)\sigma^{-2n} (\tau - \tau')}
\|  \tilde q^2  \tilde f_{\tau'} \|_{\HHq } \cr
& + \int^{\tau}_ {\tau'}\d s\, e^{(\beta ^2-\beta \tc+3\gamma/2)
\sigma^{-2n} (\tau - s)}
 \sup_{\kappa \in \real} \left| e^{-\kappa^2 (\tau -s)}
 |\kappa|^{2-\epsilon'}  \right|
 \cr & \times
  \sigma^{-2n} \sigma^{-\epsilon' n}
\| a_n\|_{{\cal C}_{\rm b}^{0,\epsilon '}}\,
  \|  \tilde f_{s}  \|_{\HHq }~,}
$$
where $\| g\|_{{\cal C}_{\rm b}^{0,\epsilon
'}}=\sup_{x\in\real}|g(x)|+\sup_{x\in\real}|(\FF^{-1}  (\tilde m \tilde{g})(x) | $
with $ \tilde m (k) =
 |1+k^2|^{\epsilon
'/2}$.
Clearly, $\| a_n\|_{{\cal C}_{\rm b}^{0,\epsilon '}}$ is finite and
using the estimate \equ{first} to bound $ \|  f_{s} \|_{\HHq } $, we get
$$
\|\tilde  q^2  \tilde f_{\tau}\|_{\HHq }
\,\leq\,
C \| (1+\tilde q^2)  \tilde f_{\tau'}\|_{\HHq }
e^{(\beta ^2-\beta \tc+1+3\gamma/2)\sigma^{-2n} (\tau-\tau')}
\sigma^{-\epsilon' n}
\max(1,(\tau-\tau')^{(3-\epsilon')/2})~.
$$
Combining these estimates completes the proof of \clm{Snn}.\QED

\REMARK
It is easy to see that additionally the following holds:
For all $ \epsilon', \alpha \in (0,1) $ there exists a $ C_{\epsilon',\alpha}
> 0 $
such that for $1>\tau >\tau '\ge0$ one has
$$
\eqalign{
\|\tilde S_{n,\tau,\tau'}\tilde  f\|_{\HHq}
\,&\le\,  C_{\epsilon',\alpha} \sigma^{-\epsilon' n}e^{-\gamma
\sigma^{-2n}(\tau -\tau ')/ 2} (\tau -\tau ')^{\alpha}
\|(1+ |\cdot|^2)^{-\alpha/2} \tilde f\|_{\HHq}~,\cr
}
$$
for all $ n \in \integer $.

\SUBSECT{apriori}{An a priori bound on the non-linear problem}We now
state and prove a priori bounds on the solution of \equ{new1}
and \equ{new3}.
Finally these solutions will be controlled by proving
inequalities
for the elements of  the following sequences.
\CLAIM{Definition}{Rn}For all $n$, we define
$$
\ru _{n} =  \|\tu_{n,1
}\|_{\HHvoid}  \quad\text{and}\quad
{\rho}_{n}^w =  \|\tw_{n,1
}\|_{\HHvoid}~.
$$
Moreover, we define
$$
\Ru_{n}\,=\,\sup_{\tau \in[\sigma^2,1]} \|\tu_{n,\tau
}\|_{\HHvoid} \quad\text{and}\quad
R_n^w\,=\,\sup_{\tau \in[\sigma^2,1]} \|\tw_{n,\tau
}\|_{\HHvoid}~.
\EQ{hyp}
$$

\CLAIM{Lemma}{apriori}For all $n \in \natural $ 
there is a constant $\eta_n>0$ such that
the following holds:
If $ \ru _{n-1} $, $ {\rho}_{n-1}^w $, and $ \sigma >0 $
are smaller than $\eta_n$,
the solutions of
\equ{new1} and \equ{new3} exist for all $ \tau \in [\sigma^2,1] $.
Moreover, we have the estimates
$$
\Ru_n
\,\le\, C \sigma^{-5/2} \ru _{n-1} + C e^{-C \sigma^{-n}}
 R_n^w +
C \sigma^{n(p-3)} (\Ru_n)^p  ~,
\EQ{after1}
$$
and
$$
R_n^w
\,\le\, C \sigma^{-5/2- \epsilon' n}{\rho}_{n-1}^w +
C \sigma^{n(p-1-\epsilon')} (\Ru_n)^{p-1} R_n^w ~,
\EQ{after2}
$$
with a constant $ C $ independent of $ \sigma $ and $ n $.

\REMARK There is no need for a detailed expression
for $ \eta = \eta_n $ since the existence of
the solutions is guaranteed if we can show
$ R^u_n < \infty $ and $ R^w_n < \infty $.
With \equ{after1} and \equ{after2} we have detailed control of these
quantities in terms of the norm of the initial conditions and $ \sigma $.

\PROOF We start with \equ{new3}. We bound the first term
of \equ{new3} by using a variant of \equ{sbound}:
First note that $ \bigl(\tilde S_{n,\tau,\sigma^2 }
\tw_{n,\sigma^2}\bigr)(\kappa)=\bigl(\tilde\LL\bigl(\tilde S_{n-1,\tau \sigma^{-2},1}
\tw_{n-1,1 }\bigr)\bigr)(\kappa)$.
Therefore,
$$
\eqalign{
\|\tilde\LL\bigl(\tilde S_{n-1,\tau \sigma^{-2},1 }
\tw_{n-1,1 }\bigr)\|_{\HHvoid}
\,&=\,\sigma^{-5/2}\|\tilde S_{n-1,\tau \sigma^{-2},1 }
\tw_{n-1,1 }\bigr)\|_{\HHvoid} \cr
\,&\le\, C\sigma^{-5/2}\sigma^{-\epsilon' n}e^{-\gamma
\sigma^{-2n}(\tau-\sigma^2 )/
2}\|\tw_{n-1,1}\|_{\HHvoid}~.\cr
}
\EQ{sboundnminus1}
$$
Therefore, we get for the first term in \equ{new3} a bound
$$
C\sigma^{-5/2}  \sigma^{- \epsilon' n}
\rho_{n-1}^w~.
\EQ{expo}
$$
For the second term in \equ{new3}, we get a bound
$$
\eqalign{
C \sigma^{n(p-3)} \int _{\sigma^2 }^\tau& \d \tau '\,
\sigma^{- \epsilon' n}e^{-\gamma \sigma^{-2n}
(\tau' -\sigma^2  )/2
 } (\Ru_n)^{p-1} R_n^w \cr
\,&\le\,
C \sigma^{n(p-3-\epsilon')}\sigma^{2n} (\Ru_n)^{p-1} R_n^w\,\le\,
C \sigma^{n(p-1-\epsilon')} (\Ru_n)^{p-1} R_n^w
~.
}
$$

We next consider \equ{new1}. The first term
is bounded by
$$
\eqalign{
\|\kappa\mapsto &
e^{-\kappa^2 (\tau -\sigma^2) }\tu_{n-1,1}(\sigma\kappa)\|_{\HHvoid}\cr
\,&\le\,
\|\kappa\mapsto
e^{-\kappa^2 (\tau-\sigma^2)
}\|_{\Cbb}\,\|\kappa\mapsto\tu_{n-1,1 } (\sigma\kappa)\|_{\HHvoid}\cr
\,&\le\, C  \sigma^{-5/2}\ru _{n-1}~,
}\EQ{new21}
$$
using \equ{Lnbound}.
Using \equ{taa} and \equ{id2}, the second term can be rewritten as
$$
\eqalign{
\sigma^{-2n}&\int_{\sigma^2 }^{\tau } \d \tau '\, e^{-\kappa^2 (\tau -\tau ')}
\bigl ((\widetilde\TT_{c\sigma^{-2n} \tau '}\ta)*(\tilde\LL^{-n} \tilde u_{n,\tau '})\bigr
)(\sigma^n\kappa) \cr
\,&=\,
\sigma^{-2n}\int_{\sigma^2 }^{\tau }\d \tau '\, e^{-\kappa^2 (\tau -\tau ')}
\bigl ((\widetilde\TT_{c\sigma^{-2n} \tau '}\ta)*\tilde u_{\sigma^{-2n}\tau '}\bigr
)(\sigma^n\kappa) \cr
\,&=\,
\sigma^{-2n}\int_{\sigma^2  }^{\tau }\d \tau '\,e^{-\beta \sigma^{-2n}
\tau'(c-\tc)} e^{-\kappa^2 (\tau -\tau ')}
\cr
&~~~\times\int \d \ell\,e^{i(\kappa-\ell) c \sigma^{-2n}\tau'}
\ta(\sigma^n \kappa -\ell-i\beta)\, \tw(\ell,\sigma^{-2n}\tau
')e^{-i\ell\tc \sigma^{-2n}\tau'} e^{-\gamma \sigma^{-2n} \tau'}~. \cr
}
$$
Using this identity, we get from the techniques leading to
\equ{tab}:
$$
\eqalign{
\sigma^{-n}&\,\|\kappa\mapsto \int_{\sigma^2}^\tau \d \tau'\,
e^{-\kappa^2(\tau-\tau')}
\bigl((\widetilde\TT_{c\sigma^{-n}\tau '}\ta_n)*\tu_{n,\tau' }\bigr)(\kappa)\|_{\HHvoid}\cr
\,&\le\,\sigma^{-n} \int_{\sigma^2  }^{\tau } \d\tau '
\|(\widetilde\TT_{c\sigma^{-n}\tau '}\ta_n)*\tu_{n,\tau' }\|_{\HHvoid}\cr
\,&\le\,\sigma^{-2n} \int_{\sigma^2  }^{\tau } \d\tau 'e^{-\beta
(c-\tc)\sigma^{-2n}\tau '}
\|\kappa\mapsto e^{-ic\sigma^{-2n}\tau '\kappa}\ta(\sigma^n
\kappa-i \beta)\|_{\HHvoid} \cr
&~~~~~~~~~~~~~~~~~~~~~\times\,\|\kappa\mapsto e^{-i\kappa\tc
\sigma^{-2n}\tau'}\tw_{n,\tau' }(\kappa)\|_{\HHvoid}
 e^{-\gamma \sigma^{-2n} \tau'}\cr
\,&\le\,C\sigma^{-2n} \int_{\sigma^2 }^{\tau } \d\tau '
(1+\tc\sigma^{-2n} \tau ')^2(1+c\sigma^{-2n} \tau ')^2  e^{-\beta
(c-\tc)\sigma^{-2n}{\tau'}} R_n^w
\cr
\,&\le\,C \sigma^{-6n}e^{-(\beta
(c-\tc)+\gamma)\sigma^{-2(n-1)}} R_n^w\,\,\leq\, \,C e^{-(\beta
(c-\tc)+\gamma) \sigma^{-n}} R_n^w~. \cr
}\EQ{bound555}
$$
For the last term in \equ{new1} we get a bound
$$
C\sigma^{n(p-3)}\int_{\sigma^2 }^\tau \d \tau ' (\Ru_n)^p\,\le\,
C\sigma^{n(p-3)} (\Ru_n)^p~.
\EQ{bound556}
$$
The proof of \clm{apriori} now follows by applying the contraction
mapping principle to \equ{new1} and \equ{new3}.
For $ \ru _{n-1} $, $ \rho_{n-1}^w $ and $\sigma>0$
 sufficiently
small the Lipschitz constant on the right hand side
of \equ{new1} and \equ{new3}
in $ \CC([\sigma^2,1],\HHuu) $
is  smaller than 1. An application of a classical fixed point argument
completes the proof of \clm{apriori}.\QED
\SUBSECT{IP0}{The iteration process}We next decompose the solution
 $\tu_{n,\tau}$ for $ \tau= 1 $
into a Gaussian part and a
remainder. Let $\tilde\psi(\kappa)\,=\,e^{-\kappa^2}$ and write
$$
\tu_{n,1 }(\kappa)\,=\,A_{n } \tilde\psi(\kappa) + \tr_{n }(\kappa)~,
$$
where
$\tr_{n }(0)=0$, and the amplitude $A_{n }$ is in $\real$.
We also define $\tPi:\HHuu\to\real$ by
$$
\tPi \tilde f= \tilde f\big|_{\kappa=0}~.
\EQ{pidef}
$$
Then \equ{new1} can be decomposed accordingly and takes the form
$$
\eqalignno{
A_{n}\,&=\,
A_{n-1} + \tPi\bigg(\int_{\sigma^2
}^{1}\d\tau'\,e^{-\kappa^2(1-\tau')}\bigg(
\sigma^{-n} (\widetilde\TT_{c \sigma^{-n}\tau' } \ta_n) *
\tu_{n,\tau'} + \sigma^{n(p-3)}\tu_{n,\tau'}^{*p}\bigg)(\kappa)\biggr)~,\cr
\tr
_{n }(\kappa)\,&=\,e^{-\kappa^2(1-\sigma^2)}
\tr_{n-1}(\sigma\kappa)\NR{an}&~~~~~~~~~~~~~~~+
\int_{\sigma^2
}^{1}\d\tau'\,e^{-\kappa^2(1-\tau')}\bigg(
\sigma^{-n} (\widetilde\TT_{c \sigma^{-n}\tau' } \ta_n) *
\tu_{n,\tau'} + \sigma^{n(p-3)}\tu_{n,\tau'}^{*p}\bigg)(\kappa)\cr
&~~~~~~~~~~~~~~~+e^{-\kappa^2 (1 -\sigma^2)} A_{n-1 } \tilde\psi(\sigma\kappa)-
A_{n}\tilde\psi(\kappa)~.\NR{rn}
}
$$
Then we define $ \rr_n = \| \tr_{n } \| $ and so
$ \ru _n \,\leq\, C (|A_n| + \rr_n) $.
Our main
estimate is now
\CLAIM{Proposition}{decay}There is a constant $C>0$ such that
for $ \sigma > 0 $ sufficiently small
the solution $\tu$ of \equ{uequ} satisfies for all $n \in \natural $:
$$
\eqalignno{
|A_{n}-A_{n-1}|\,&\le\,C
e^{-C \sigma^{-n}} R_{n}^w +C\sigma
^{n(p-3)}(\Ru_{n})^p~,\NR{a1}
\rr_n
\,&\le\, \rr_{n-1}/2 + C e^{-C \sigma^{-n}} R_{n}^w +C\sigma
^{n(p-3)}(\Ru_{n})^p~, \cr
\rho_n^w
\,&\le\, C e^{- C \sigma^{-2n}} \rho_{n-1}^w +
C\sigma^{n (p-1-\epsilon')
}(\Ru_{n})^{p-1} R_n^w~.\NR{a2}
}
$$

\PROOF We begin by bounding the difference $A_{n}-A_{n-1}$  using \equ{an}.
Observe that since we work in $\HHuu$, we have
$$
|\tPi\tilde  f|\,\le\,C \|\tilde f\|_{\HHvoid}~,
\EQ{Pibound}
$$
with $C$ independent of $\delta $.
Thus, it suffices
to bound the norm of the integral in \equ{an}.
The first term in \equ{an} is the one containing the translated
term $\ta_n$ and was already bounded in \equ{bound555} while the
second was bounded in \equ{bound556}.
Combining these bounds with \equ{Pibound}, we find
\equ{a1}.

We next bound $\tr_{n}$ in terms of $\tr_{n-1 }$, using
\equ{rn}. The
first term is the one where the projection is crucial:
For  $\sigma>0 $ sufficiently small,
$\tilde f\in\HHuu$ with $\tilde f(0)=0$ one has
$$
\| \kappa \mapsto e^{-\kappa^2  (1-\sigma^2)} \tilde f(\sigma
\kappa)\|_{\HHvoid}
\,\le\,  \|\tilde f\|_{\HHvoid}/2~.
\EQ{crucial}
$$
Indeed, writing out the definition \equ{hhuunorm} of $\HHuu$, one
gets for the term with
$j=\ell=0$:
$$
\eqalign{
\int \d \kappa\,
e^{-2\kappa^2 (1-\sigma^2)} |\tilde f(\sigma\kappa)|^2\,&=\,
\sigma^{-1}\int \d (\sigma\kappa)\,
e^{-2\kappa^2 (1-\sigma^2)}(\sigma \kappa)^2
\left|{\tilde f(\sigma\kappa)-\tilde f(0)\over \sigma\kappa}\right|^2~.
}
$$
Clearly, a bound of the type of \equ{crucial} follows for this term by
the assumptions on $\tilde f$. The derivatives are handled similarly, except
that there is no need to divide and multiply by powers of
$\sigma\kappa$ since each derivative produces a factor $\sigma$.

We now bound the other terms in \equ{rn}. The first term is bounded using
\equ{crucial} and yields a bound (in $\HHuu$) of
$$
 \ru_{n-1}/2~.
\EQ{bound1}
$$
The second and third terms have been bounded in \equ{bound555} and
\equ{bound556}:
$$
C e^{-C \sigma^{-n}} R_n^w +C\sigma
^{n(p-3)}(\Ru_n)^p~.
\EQ{bound67}
$$
Finally, the last term in \equ{rn} can be written as
$$
\tilde X_{n}\,\equiv\, A_{n-1} (e^{-\kappa^2(1-\sigma^2)}
e^{-\kappa^2\sigma^2}-e^{- \kappa^2})
+(A_{n-1 }-A_{n})e^{-\kappa^2}~.
$$
The first expression vanishes and we get a
bound (in
$\HHuu$):
$$
\|\tilde X_{n }\|_{\HHvoid}\,\le\,C e^{-C \sigma^{-n}} R_{n}^w +C\sigma
^{n(p-3)}(\Ru_n)^p~.
\EQ{b7}
$$
Collecting the bounds \equ{bound1}--\equ{b7}, the assertion \equ{a2}
for $\tr_{n }$ follows.
Finally, the bounds on $\rho_{n }^w$ follow as those in
\clm{apriori}.
The proof of \clm{decay} is complete.
\QED
\LIKEREMARK{Proof of \clm{main1}}The proof is an induction
argument, using repeatedly
the above estimates.
Again we write $ C $ for (positive) constants which can be chosen independent
of $ \sigma $ and $ n $.
Assume that $ R = \sup_{n \in\natural } \Ru_n < \infty $ exists.
{}From \clm{apriori}
we observe for $ \sigma > 0 $ sufficiently small
$$
\eqalign{
R_n^w  \,&\,\leq\, \, {{C \sigma^{-5/2 - n \epsilon'} \rho_{n-1}^w}\over
{1- C \sigma^{n(p-1-\epsilon')} R^{p-1}}}
\,\leq\, C\, \sigma^{-5/2 - n \epsilon'} \rho_{n-1}^w ~,\cr
\Ru_n  \,&\,\leq\, \,
{C \sigma^{-5/2} \ru _{n-1} + C e^{-C \sigma^{-n}}
R_n^w\over
         1- C \sigma^{n(p-3)} R^{p-1} } \cr
&\,\leq\, \, C \sigma^{-5/2} \ru_{n-1} + C e^{-C \sigma^{-n}}
\rho_{n-1}^w~,
}
\EQ{pp}
$$
with a constant $ C $ which can be chosen independent
of $ R $.
Using \clm{decay} we find
$$
\eqalignno{
|A_{n}-A_{n-1}|\,&\le\,C
e^{-C \sigma^{-n}} \rho_{n-1}^w +C\sigma
^{n(p-3)} \sigma^{-5/2} \ru _{n-1}~, \cr
\rr_n
\,&\le\, \rr_{n-1}/2 + C e^{-C \sigma^{-n}} \rho_{n-1}^w +C\sigma
^{n(p-3)}\sigma^{-5/2} \ru _{n-1}~, \cr
\ru _n \,&\leq\, C (|A_n| + \rr_n) ~,\cr
\rho_n^w\,&\le\,  C e^{- C \sigma^{-2n}} \rho_{n-1}^w +
C\,\sigma^{n (p-1-\epsilon')} \sigma^{-5/2 - n \epsilon'} \rho_{n-1}^w~.\cr
}
$$
Therefore, we can choose $ \sigma >0 $ so small that for
$ n > 3 $:
(recall $ p>3 $ and $ p \in \natural $)
$$
\eqalignno{
|A_{n}-A_{n-1}|\,&\le\,
\rho_{n-1}^w/10 +\sigma
^{n-3} (|A_{n-1}|+ \rr _{n-1})~, \cr
\rr_n
\,&\le\, 3 \rr_{n-1}/4 + \rho_{n-1}^w/10
+\sigma^{n-3} |A_{n-1}|
~, \cr
\rho_n^w\,&\le\,  \rho_{n-1}^w/10~ .\cr
}
$$
Thus, the
sequence of $A_n$ converges geometrically to a finite limit
$A_*$. Furthermore,
we find that
$ \lim_{n \rightarrow \infty} \rr_n = 0 $, and
$ \lim_{n \rightarrow \infty} \rho_n^w = 0 $.
Since the quantities
$ |A_n | $, $ \rr_n $, $  \rho_n^w $
increase only for at most three steps the term $ C R^{p-1} $
in \equ{pp} stays less than $ 1/2 $ if we choose
$ |A_1 | $, $ \rr_1 $, $  \rho_1^w  = \OO( \sigma^m ) $,
for an $ m >0 $
sufficiently large.
We then deduce from \equ{pp} the existence of a finite constant
$ R = \sup_{n \in\natural } \Ru_n$.
Finally,
the scaling of $\tilde{w}_{n,\tau}$ implies the exponential decay
of $\tilde{w}_t$.
The proof of \clm{main1} is complete.\QED
\vfill
\eject

\SECTIONNONR{Part II. The Swift-Hohenberg equation}
\global\advance\SECTIONcount by -1
\vskip-1truecm
\SECT{Bloch waves}{Bloch waves}Since the problem we consider takes
place in a setting with a {\em periodic} background provided by the
stationary solution of the Swift-Hohenberg, it is natural to
work with the Bloch representation of the functions.
For additional informations see [RS72].

The starting point of Bloch wave analysis
in case of a $ 2 \pi $--periodic underlying pattern
is the following relation
$$
\eqalign{
u(x) \,&=\, \int \d k\, e^{ikx} \tilde{u}(k)
    \,=\,\sum_{n \in \integer} \int_{-1/2}^{1/2} \d\ell\,e^{i(n+\ell)x}
     \tilde{u}(n+\ell)  \cr
     \,&=\,  \int_{-1/2}^{1/2} \d \ell\,\sum_{n\in \integer} e^{i(n+\ell)x}
     \tilde{u}(n+\ell )
     \,=\,  \int_{-1/2}^{1/2} \d\ell\,e^{i\ell x} \ub(\ell,x)~,\cr
}
\EQ{abbi}
$$
where we define
$$
\bigl(\JJ u\bigr)(\ell,x)\,\equiv\, \ub(\ell,x)\,=\,\sum_{n \in
\integer}e^{inx} \tu(n+\ell )~.
\EQ{rumme}
$$
The operator $\JJ$ will play a r\^ole analogous to that played by the
Fourier transform $\FF$ for the simplified problem of Part I.
We will use analogous notation:
\LIKEREMARK{Notation}If $f$ denotes a function, then $\hat f$ is
defined by $\hat f=\JJ f$, and if $\AA$ is an operator, then $\hat
\AA$ is defined by $\hat \AA= \JJ \AA \JJ^{-1}$.

Note that
$$
\int_\real \d x\, | u(x) |^2 \,=\, 2\pi \int_{-1/2}^{1/2} \d\ell\int
_0^{2\pi}\d x\, |\ub(\ell,x)|^2~.
\EQ{rg0}
$$
This is easily seen from Parseval's identity:
$$
\eqalign{
\int_{\real} \d x \,|u(x)|^2
\,&=\, 2 \pi  \int_{\real} \d k|\tilde{u}(k)|^2   \cr
     \,&=\, 2\pi  \sum_{n \in \integer} \int_{-1/2}^{1/2}
    \d \ell\,  |\tilde{u}(n+\ell )|^2  \cr
     \,&=\,  2 \pi \int_{-1/2}^{1/2}\d \ell
     \sum_{n \in \integer} |\tilde{u}(n+\ell )|^2   \cr
     \,&=\,  2 \pi  \int_{1/2}^{1/2} \d\ell
      \int_0^{2 \pi}\d x\, |\ub(\ell,x)|^2 ~.\cr
}
$$
The sum and the integral can be interchanged
in \equ{abbi} due to Fubini's theorem when
$ u$ is in the Schwartz space  $\SS $.

We shall use frequently
the following fundamental properties (which follow at once from \equ{rumme}):
$$
\eqalign{
\ub (\ell,x)\,&=\,e^{i  x} \ub (\ell+1,x)~, \cr
\ub (\ell,x)\,&=\,\ub (\ell,x+2\pi)~, \cr
\ub (\ell,x)\,&=\,\overline{\ub}(-\ell,x) \ \text{for \ real--valued } u~.\cr
}
\EQ{rg1}
$$
Multiplication in position space corresponds to a modified convolution
operation for the Bloch-functions:
$$
\bigl(\widehat{u\cdot v}\bigr )(\ell,x)\,=\,
\int_{-1/2}^{1/2} \d\ell'\,
\hat u(\ell-\ell',x)\hv (\ell',x)\,\equiv\,\bigl(\hat u\starb
\hv \bigr)(\ell,x)~.
$$
This follows from \equ{rg1} and the identities:
$$
\eqalign{
\bigl(\widehat{u\cdot v}\bigr )(\ell,x)\,&=\,
\sum_{m\in\integer} \int_\real \d k\,
\tilde u(\ell +m -k)\tilde v(k) e^{i mx}\cr
\,&=\,
\int_{-1/2}^{1/2} \d\ell'\,
\sum_{m,n\in\integer}
\tilde u(\ell +m -\ell'-n)\tilde v(\ell'+n) e^{i (m-n)x}e^{inx}~.\cr
}
$$
Recalling the norm
$$
\| f\|_{\Huu}\,=\,\left (\sum_{j,m=0}^2\delta ^{2(m+j)}\int \dx
|\partial_x^m  f(x)|^2 x^{2j} \right )^{1/ 2} 
$$
we now introduce
$$
\|\hat f\|\,\equiv\,\|\hat f\|_{\HHbuu}\,=\,
\left (\sum_{j,m=0}^2\delta ^{2(m+j)}\int_{1/2}^{1/2}\dell  \int_0^{2\pi} \dx
|\partial_x^j
\partial_\ell^m \hat f(\ell,x)|^2 \right )^{1/ 2}~.
$$
We get from Parseval's equality
$$
C^{-1}\|u\|_{\Huu}\,\le\,  \|\hat u \|_{\HHbuu}\,\le\,
C\|u\|_{\Huu}~,
$$
for some $C$ independent of $\delta \in (0,1) $.
Similarly, in analogy to \equ{fg}, we also have
$$
\eqalignno{
\|\widehat{u\,v}\|_{\HHbuu}\,=\,\|\hat u\starb \hv \|_{\HHbuu}
\,&\le\,
C\|\hat u\|_{\HHbuu}\,\,\|\hv \|_{\HHbuu}~,\NR{bl1}
\|\bigl(\widehat{u\,v}\bigr )(\cdot-i\beta,\cdot )\|_{\HHbuu}\,
=\,\|\bigl(\hat u\starb \hv \bigr)(\cdot-i\beta,\cdot )\|_{\HHbuu}
\,&\le\,
C\|\hat u\|_{\HHbuu}\,\,\|\hv (\cdot-i\beta,\cdot )\|_{\HHbuu}~.\NR{bl2}
}
$$
Finally, suppose $f$ is a function in $\Cbb$ (see \equ{cbbdef} for the
definition): Then,
$$
\eqalignno{
\|\widehat{f\,v}\|_{\HHbuu}\,=\,\|\hat f\starb \hv \|_{\HHbuu}
\,&\le\,
C\| f\|_{\Cbb}\,\,\|\hv \|_{\HHbuu}~,\NR{bl3}
\|\bigl(\hat f\starb \hv \bigr)(\cdot-i\beta,\cdot )\|_{\HHbuu}
\,&\le\,
C\| f\|_{\Cbb}\,\,\|\hv (\cdot-i\beta,\cdot )\|_{\HHbuu}~.\NR{bl4}
}
$$
Thus, apart from notational differences, we can work in the Bloch
spaces with much the same bounds as in the spaces used for the model
problem of the previous sections.

\SECT{fulllin}{The linearized problem}We discuss here again the
behavior of the linearized problem as in \sec{linear}, but now for the
Swift-Hohenberg equation.
The discussion will again be split in an aspect behind the front and
one ahead of the front.
In \sec{linear}, the behavior of the
problem in the bulk behind the
traveling front was diffusive by construction, and the only
difficulty was to understand the r\^ole of the decay of $a$ to 0 (as
$e^{-\beta |x|}$) as $x\to-\infty $.
For the problem of the Swift-Hohenberg equation, the situation is
similar, leading again to diffusive behavior. However, this
observation is not obvious. Therefore, the
first problem consists in showing the diffusive behavior.
In order to obtain
optimal results for the analysis ahead of the front, \ie for the variable
in the weighted representation, we use our approximate knowledge
of the shape of the front.

\SUBSECT{unweightsh}{The unweighted representation}In analogy with the
simplified
example, the linearized problem would be  now
$$
\partial_t v  \,=\, \MM v  +\MMi v ~,
\EQ{m61}
$$
where $\MM$ and $\MMi$ have been defined in Eqs.\equ{13a} and
\equ{13b}.
By the analysis for the model problem we expect that the term
$ \MMi v $ will be irrelevant for the dynamics in the bulk
with some exponential rate.
Therefore, it will be considered in the sequel together with
the non-linear terms. As a consequence,
the linear equation dominating the behavior behind the
front is given by
$$
\partial_t v\,=\, \MM v~.
\EQ{mmlin}
$$
We recall those features
of the proof of diffusive
stability of [Schn96, Schn98] which are relevant to the study of \equ{mmlin}.

In order to do this, we need to localize the spectrum of $\MM$. Since
this is well-documented, we just summarize the results.
As the linearized problem has periodic coefficients, the operator
$\hat \MM=\JJ\MM\JJ^{-1}$ equals a direct integral $\int^\oplus \d\ell
\,\MM_\ell$, where
each $\MM_\ell$ acts on the subspace with fixed quasi-momentum $\ell$
in $\HHbuu$.
The
eigenfunctions of $\MM_\ell$ are given by Bloch waves of the form $e^{i\ell
x} w_{\ell,n}$ with $2\pi$-periodic $w_{\ell,n}$. The index
$n \in \natural $ counts
various eigenvalues for fixed $\ell$. For each $\ell\in\real$ (or
rather in the Brillouin zone $[-\HALF,\HALF]$) they are solutions of the
eigenvalue equation
$$
\bigl(\MM_\ell w_\ell\bigr )(x)\,\equiv\,-\bigl(1+(i\ell+\partial_x)^2\bigr)^2 w_\ell(x) +\epsilon ^2 w_\ell(x)
-3 \Ustar ^2(x)w_\ell(x)\,=\,\mu_\ell w_\ell(x)~.
$$
The spectrum takes the familiar form of a curve $\mu_1(\ell)$ with an
expansion
$$
\mu_1(\ell)\,=\,-c_1\ell^2+\OO(\ell^3)~,
$$
and $c_1>0$ and the remainder of the spectrum negative and
bounded away from 0.
The eigenfunction associated with $\mu_1(0)$ is $\partial_x \Ustar (x)$,
reflecting the translation invariance of the original problem \equ{sh}.
There is an $\ell_0>0$ such that for fixed
$\ell\in(-\ell_0,\ell_0)$  the eigenfunction $\phi_\ell(x)=
w_{\ell,1}(x)$ of the main
branch $\mu_1(\ell)$ is well defined (and a continuation of
$\partial_x \Ustar (x)$)
as $\ell$ is varied away from 0.
Corresponding to this we define the central projections $\Pc(\ell)$ by
$$
\Pc(\ell) f\,=\, \langle \bar\phi_\ell,f\rangle \phi_\ell~,
$$
where $\langle\cdot,\cdot\rangle$ is the scalar product in
$\L^2([0,2\pi])$ and $  \bar\phi_\ell $ the associated
eigenfunction of the adjoint problem.
We will need a smooth version of the projection in $\HHbuu$. We fix once
and for all a non-negative smooth cutoff function $\chi$ with support in
$[-\ell_0/2,\ell_0/2]$ which equals 1 on $[-\ell_0/4,\ell_0/4]$. Then
we define the operators $\Ec$ and $\Es$ by:
$$
\Ec(\ell)\,=\,\chi(\ell)\Pc(\ell)~,\quad
\Es(\ell)\,=\,{\bf 1}(\ell)-\Ec(\ell)~.
$$
It will be useful to define auxiliary ``mode filters'' $\Ech$ and
$\Esh$
by
$$
\Ech(\ell)\,=\,\chi(\ell/2)\Pc(\ell)~,\quad
\Esh(\ell)\,=\,{\bf 1}(\ell)-\chi(2 \ell)\Pc(\ell)~.
$$
These definitions are made in such a way that
$$
\Ech \Ec\,=\,\Ec~,\quad
\Esh \Es \,=\,\Es~,
$$
which will be used to replace the (missing) projection property of
$\Ec$ and $\Es$.

We next extend the definitions \equ{mndef}
of \sec{Renormalization} to the Bloch spaces. To
avoid cumbersome notation, we shall use mostly the same symbols as in
that section.
Thus, with $\sigma<1$ as before, we let now
$$
\bigl(\hLL \hat u\bigr)(\kappa,x)\,=\,\hat u(\sigma \kappa,x)~.
$$
Note that here, and elsewhere, the scaling does not act on the $x$
variable, only on the quasi-momentum $\kappa$.
The novelty of renormalization in Bloch space here is that
since the integration region over the $\ell$
variable is finite it will change with the scaling.
Therefore, we introduce (for fixed $\delta >0$),
$$
\KK_{\sigma,\rho}\,=\,\{
\hat u~|~
\|\hat u\|_{\KK_{\sigma,\rho}}<\infty ~,
\}
\EQ{kkdef}
$$
where
$$
\|\hat u\|^2_{\KK_{\sigma,\rho}}\,\equiv\,
\sum_{n,n'=0}^2 \int_{-1/(2\sigma)}^{1/(2\sigma)}\d \ell
\int_{0}^{2\pi}\d x\,\, \delta^{2(n+n')}
|\partial_\ell^n \partial_x^{n'}
\hat u(\ell,x)|^2 (1+\ell^2)^\rho~.
$$
For technical reasons we introduced a weight in the Bloch variable
$ \ell $.
We will always write $\KK_\sigma$ instead of $\KK_{\sigma,1}$.
Note that $\JJ$, as defined in \equ{rumme} is an isomorphism
between the space $\Huu$ and the space $\Zuu$ by \equ{rg0} and the
definition \equ{kkdef}.

Consider again the eigenfunctions
$\phi_\ell(x)$. The function
$$
\hv _t(\ell,x)\,=\,e^{\mu_1(\ell)t}\phi_\ell(x)~,
$$
solves the equation
$$
\partial_t \hv _t(\ell,\cdot)\,=\,\MM_\ell (\hv _t(\ell,\cdot))~.
$$
Because of the nature of the spectrum $\mu_1(\ell)$, this solution
satisfies
$$
\hv _t (\ell t^{-1/2},x)\,=\,
e^{-c_1\ell^2 } \hv _0(0,x)+\OO(t^{-1/2})~.
$$
Using this observation and the fact that the $ \Es$-part is
exponentially damped, the result will be

\CLAIM{Proposition}{problem1}The solution $\widehat V_t$ of the problem
\equ{mmlin}
with initial data $\widehat V_0$ satisfies:
$$
\|(\ell,x)\mapsto \widehat V_t(\ell t^{-1/2},x)-e^{-c_1 \ell^2 }
\Pc(0) \widehat
V_0(0,x)\|_{\KK_{1/\kern-0.15em\sqrt{t}}}\,\le\,
{C\over t^{1/2}}\|\widehat V _0\|_{\HHbuu}~,
\EQ{p2bound}
$$
for a constant $ C > 0 $ and all $ t \geq 1 $.
Moreover, there is a constant $ {\gamma}_- > 0 $ such that
$$
\eqalignno{
\|(\ell,x)\mapsto \bigl(\Es \ \widehat V_t\bigr)
(\ell t^{-1/2},x)\|_{\KK_{1/\kern-0.15em\sqrt{t}}}\,&\le\,
C {e^{-\gamma_- t}} \| \widehat V _0\|_{\HHbuu}~,\NR{p1.2}
}
$$
for all $ t \geq 1 $.

\SUBSECT{weightsh}{The weighted representation}The weighted
representation will be obtained by translating the
effect of the
transformation $\WW$ defined in \equ{WWdef} to the language of the
Bloch waves.
In accordance with our notational conventions, we set
$$
\widehat \WW \,=\,
\JJ\WW\JJ^{-1}~,
$$
and we get now, in analogy to \equ{WWhatdef},
$$
\bigl(\widehat \WW  \hat f \bigr)(\ell,x)
\,=\,
e^{i\tc(\ell+i\beta )t}\hat f (\ell+i\beta ,x+\tc t)~.
$$
The equation \equ{m61}, expressed in terms of $\widehat \WW \hv$,
then takes the form
$$
\partial_t \bigl(\widehat \WW \hv  \bigr)\,=\,\widehat \MM_{\beta ,\tc
t}\bigl(\widehat \WW \hv  \bigr)+ \widehat {\MM}_{{\bf i},\beta ,\tc
t}\bigl(\widehat \WW \hv
\bigr)~,
\EQ{mmlin2}
$$
with
$$
\eqalign{
\bigl(\widehat \MM_{\beta ,\tc t }\hat  f\bigr)(\ell,x)\,&=\, \bigl(\hat
L_{i\beta }\hat f\bigr)(\ell,x)-3\Ustar ^2(x) \hat f(\ell,x)
+ \tc (i(\ell+i\beta
)+\partial_x)\hat f(\ell,x)~,\cr
\bigl(\widehat \MM_{{\bf i},\beta ,\tc t }\hat  f\bigr)(\ell,x)\,&=\,
-6\Ustar (x)\bigl(\hKc\starb
\hat f\bigr)(\ell,x)  -3(\hKc\starb\hKc\starb \hat f )(\ell,x)~.\cr
}
$$
Some explanations are in order: $\hat L_{i\beta }$ is the operator
$-(1+(\partial_x + i\ell -\beta)^2)^2 +\epsilon ^2$.
The functions $\Ustar $ are just multiplications in the Bloch representation
because they are periodic. More precisely, one has $\widehat \Ustar (\ell,x)=
\Ustar (x)\delta (\ell)$ in the sense of distributions.
The functions $\hKc$ are derived from $\Kc$ of Eq.\equ{kcdef} and
are seen to be given by
$$
\hKc(\ell,x) \,\equiv\, \bigl(\JJ \Kc\bigr)(\ell,x)\,=\,e^{-i\ell c t}
\widehat F_c(\ell,x-ct,x)-\Ustar (x)\delta (\ell)~,
$$
where the Bloch transform is taken in the first (non-periodic)
variable of $F_c$.

In order to obtain
optimal results for the analysis ahead of the front, \ie for the variable
in the weighted representation, we the recall some facts from the construction
[CE86, EW91]
of the fronts.

For small $\epsilon > 0$ the bifurcating solutions $u$ of the Swift-Hohenberg
equation can be approximated by
$$
\tilde\psi (x,t,\epsilon) \,=\, \epsilon A(\epsilon
x,\epsilon^2t)e^{ix} + \text{c.c.}~,
$$ 
up to an error $\OO(\epsilon^2)$, where $A$
satisfies the Ginzburg-Landau equation
$$
\partial_TA \,=\, 4\partial^2_{X}A + A - 3A|A|^2~,
$$
with ${X} \in \real$, $T \ge 0$ and $A({X},T) \in \complex $.
See [CE90b, vH91, KSM92, Schn94].
This equation possesses a real-valued
front $A_{\rm f}(X,T) = B(X-c_B T ) $, where
$\xi\mapsto B(\xi)$ satisfies the ordinary differential equation
$$
4B'' + c_B B' + B - 3B|B|^2 \,=\, 0~.
$$
For $|c_B | \geq 4 $ the real--valued  fronts of this equation are monotonic.
These fronts and the trivial solution  $ A = 0 $ can be stabilized
by introducing a weight $ e^{\betaA x} $ satisfying
the stability condition
$$
\varrho_A(c_B,\betaA ) = 4 \betaA ^2 - \betaA  c_B + 1 < 0 ~,
$$
see [BK92].

\REMARK Since $ B(\xi) $ converges at a faster rate to $1/\sqrt{3} $ for
$ \xi \rightarrow - \infty $ than to $ 0 $ for
$ \xi \rightarrow \infty $
there will be  no
additional restriction such as \equ{a4} on $ \betaA $.

\REMARK
Our result will be optimal in the sense that
each modulated front $ F_c $ which
corresponds to a front of the associated
amplitude equation satisfying $ \varrho_A(c_B,\betaA ) < 0 $
is stable.
The connection between the quantities of the
Ginzburg-Landau equation and the associated Swift-Hohenberg equation
is as follows.
We have
$ c = \epsilon c_B + \OO(\epsilon^2) $,
and $ \beta = \epsilon \betaA + \OO(\epsilon^2) $.

In order to prove this remark we write the modulated front $ F_c $
as defined in \equ{fcqa} as a sum of the Ginzburg-Landau part and a remainder
$$
F_c(\xi,x) \,=\,
2 \epsilon B(\epsilon \xi) \cos(x) + \epsilon^2 F_{\rm r}(\xi,x)~,
$$
where $ F_{\rm r} $ satisfies
$$
\sup_{y \in \real} \| F_{\rm r}(\cdot + y,\cdot) \|_{\Cbb} \,\leq\,
C~,
$$
for a constant $ C $ independent of $ \epsilon \in (0,1)$
and $ \delta \in (0,1) $.
Then we consider \equ{mmlin2} which we write without decomposition as
$$
\partial_t \widehat W \,=\,
\bigl(
\widehat L_{i\beta }\widehat W \bigr)-3 (\widehat {\tau_{ct} F}) \starb
(\widehat {\tau_{ct} F}) \starb
 \widehat W + \tc (i(\ell+i\beta
)+\partial_x)  \widehat W~.
\EQ{a1shh}
$$
In order to control these solutions
we use that
the linearized system \equ{mmlin2} evolves in such a way that during
times of order $ \OO(1/\epsilon^2) $ it can be approximated
by
the associated linearized
Ginzburg-Landau equation
$$
\partial_\tau  A = 4 (\partial_X-\betaA )^2 A + c_B
(\partial_X-\betaA ) A+ A - B^2(2 A + \overline{A})~.
\EQ{GLlin}
$$
\CLAIM{Theorem}{pseu}For all $C_0 >0 $, and $\tau _1>0$
there exist positive constants
$\epsilon_0 $,
$ C_1 $,
$C_2$, and $\tau _0$ such that for all $\epsilon
\in (0,\epsilon_0]$
the following is true:
For all initial conditions
$\widehat W_0$ with $
\|\widehat W _0\|_{\HHbuu}\,\leq\, C_0 \epsilon $
there are a
solution $\widehat W_t $ of \equ{a1shh}
and a solution $ A_\tau  $ of \equ{GLlin}
with $ \| A_0 \|_{\HHuu} \leq C_1 $
such that
the function $A_\tau $
approximates $\widehat W_t$ in the sense that
$$
\|\widehat W_t- \epsilon \JJ (A_{\epsilon^2t-\tau _0}(x) e^{ix} + {\rm c.c.})
\|_{\HHbuu} \,\leq\, C_2 \epsilon^{2}~,
$$
for all $ t\in [\tau_0/\epsilon^2,(\tau_0 + \tau_1)/\epsilon^2]$.
Here $\JJ $ again denotes the map of Eq.\equ{rumme} from a function
$f$ of $x$ to
its Bloch representation $\hat f(\ell,x)$.

\PROOF The proof of this is very similar to the case of the (non-linear)
Swift-Hohenberg equation which
was discussed in the literature [CE90b, vH91, KSM92, Schn94]. Our (linear) 
problem is in
fact easier and the proof is left to the reader.\QED

For the system \equ{GLlin} we have the estimate [BK92]
$$
\| A_\tau  \|_{\H^2_{2,\delta}} \,\leq\, C e^{\varrho_A(c_B,\betaA,\delta
) \tau }
\| A_0 \|_{\H^2_{2,\delta}}~,
$$
with $ \lim_{\delta \rightarrow 0}
\varrho_A(c_B,\betaA,\delta) = \varrho_A(c_B,\betaA) $.
The deviation of  $ \varrho_A(c_B,\betaA,\delta)$ from $ 
\varrho_A(c_B,\betaA) $ comes again from the derivatives of $ B $ 
and from the polynomial weight in the norm  $\H^2_{2,\delta}$.
As a consequence of this estimate and of \clm{pseu} we conclude
that
$$
\| \widehat W_t \|_{\HHuu} \,\leq\,
C e^{\varrho(c,\beta,\epsilon,\delta) (t-t')}
\| \widehat W_{t'} \|_{\HHuu}~,
\EQ{varrho}
$$
for a constant $ C $ and a coefficient
$ \varrho = \varrho(c,\beta,\epsilon,\delta) $.
We can (and will) choose this constant $\varrho $ in such a way that
(for $\epsilon \to 0$):
$$
\varrho(c,\beta,\epsilon,\delta) \,=\, \epsilon^2 
(\varrho_A(c_B,\betaA,\delta ) + o(1) )~.
\EQ{sccompare}
$$
We define
$ \varrho(c,\beta,\epsilon) = \lim_{\delta \rightarrow 0}
 \varrho(c,\beta,\epsilon,\delta)$.

\REMARK
The choice  of a sufficiently small $ \delta>0  $ 
and $ \epsilon > 0 $ 
will allow us
 to
prove the stability of all fronts which are predicted to be stable
by the associated amplitude equation
since $ \lim_{(\epsilon,\delta) \rightarrow 0} 
\epsilon^{-2} \varrho(c,\beta,\epsilon,\delta) = \varrho_A(c_B,\betaA) $.

In the following 
we consider a modulated front
with velocity  $ c $ and
a given (sufficiently small) bifurcation parameter $ \epsilon >0 $
for which there are a $ \beta $ 
and a $ \tc \in (0,c) $
which satisfy:
$$
\varrho(\tc,\beta,\epsilon) = -2 \gamma < 0 ~.
\EQ{sc}
$$

\CLAIM{Proposition}{problem23}Suppose that the above stability
condition \equ{sc} is satisfied. Then there is a $ \delta \in (0,1] $
such that: 
There is a $C<\infty $ for which the
functions $\widehat
W_t=\widehat \WW  \widehat V_t$
obey the bounds
$$
\| \widehat W_t \|_{\HHbuu}\,\le\,
C e^{-3 \gamma (t-s)/2 } \|\widehat W_s\|_{\HHbuu}~.
\EQ{p3bound}
$$

As in the previous sections this result will have to be improved
for the non-linear problem. Therefore, we skip at this point the proof,
and will only deal with the improved version later.

Thus, the linear problems \equ{mmlin} and \equ{mmlin2} are the analogs of
\equ{hatulin} and \equ{hatwlin} and can be studied pretty much as in
the case of the
simplified problem, yielding  inequalities  similar to \equ{res1} and
\equ{res2}.

\SECT{renorm}{The renormalization process for the full problem}We
assume throughout this section
that the stability condition \equ{sc} is
satisfied. We
prove here our main

\CLAIM{Theorem}{main2}There are a $ \delta > 0 $ and
positive
constants $R$ and $C$ such that the following
holds:
Assume $\|v_0\|_{\Huu}+\|M
_\beta v_0\|_{\Huu} \le R$ and denote by $v_t$ the solution of
\equ{form1} with
initial condition $v_0$. Let $\tpsi (\ell)=\exp({-c_1  \ell^2})$.
There is a constant $A_*=A_*(v_0)$ such that
the rescaled solution $ \hat{v}_t^\r(\ell,x) = \hv _t ( \ell t^{-1/2},x ) $
satisfies
$$
\|\hat{v}^\r_t-A_* \tpsi \partial_x \Ustar
\|_{\KK_{1/\kern-0.15em\sqrt{t}}}\,\le\,
{CR\over (t+1)^{1/4}}~.
\EQ{mainineqsh}
$$
Furthermore,
$$
\|\hw _t\|_{\KK_{1/\kern-0.15em\sqrt{t}}}\,=\,\| \widehat \WW \hat v _t
\|_{\KK_{1/\kern-0.15em\sqrt{t}}}\,\le\,CR
e^{-\gamma t}~.
\EQ{mainineqsh22}
$$

\LIKEREMARK{Remarks}\item{$\bullet$}The inequality \equ{mainineqsh}
really says that the difference
$$
\hv _t (\ell t^{-1/2},x) -A_* e^{-c_1\ell^2} \partial _x\Ustar (x)
$$
is small, where $\Ustar $ is the periodic solution (see Eq.\equ{periodic})
of the Swift-Hohenberg equation.
Expressed in the laboratory frame, this means that {\em an initial
perturbation $v_0(x)$ will go to 0 like}
$$
v_t(x)\,\approx\, A_*(v_0)\sqrt{\pi\over c_1 t} \exp({-x^2\over 4c_1t})\,\partial _xU_* (x)~,
$$
when $t\to\infty $, uniformly for $ x \in \real $.
See [Schn96]. In particular, this means that near the extrema of
$U_*$ the convergence is faster than $\OO(t^{-1/2})$ since at those
points $\partial _xU_*$ vanishes.
\item{$\bullet$}The inequality \equ{mainineqsh22} gives some more
precise bound on the growth of a perturbation ahead of the front,
because it says that this perturbation decays exponentially in the
weighted norm. More explicitly, we have at least a bound
$$
|v_t(x+ct)| \,\le\, C e^{\beta x - \gamma' t}~,
$$
with $\gamma'$ slightly smaller than $\gamma$
\item{$\bullet$}The decay $ (t+1)^{-1/4} $ in \equ{mainineqsh} can be improved easily to
$ (t+1)^{-1/2+\epsilon } $
for any $ \epsilon  > 0 $.
We have chosen $ \epsilon  =1/4 $ to keep the notation at a reasonable level.

\PROOF As we explained before, the proof is similar to the one in \sec{linear}
except that now the
function behind the front is split into  a
diffusive part $ \vc  $ and into an exponentially damped part $ \vs
$, and correspondingly there will be a few more equations.

In Bloch space the initial conditions satisfy
$ \| \hv _0 \|_{\HHbuu} + \| \hv _0(\cdot - i \beta,\cdot) \|_{\HHbuu}
\,\leq\, R $.
The system for the variables $ \vc $ and $ \vs $
with initial conditions $ \vc|_{t=0}  = \Ec \hv |_{t=0} $, $ \vs|_{t=0} =
\Es \hv |_{t=0} $, and for the variable
$ \hw  = \widehat \WW \hat v $
with initial conditions
$ \hw|_{t=0} = \widehat {\cal W}_{\beta,0} \hat{v}|_{t=0} $
is given in Bloch space by
$$
\eqalign{
\partial_t \vc   \,&=\,   \hMM   \vc   + \Ec  \hat{\cal H}  (\vc ,\vs )
+\Ec   \hNN (\vc ,\vs )~, \cr
\partial_t \vs   \,&=\,   \hMM  \vs  + \Es  \hat{\cal H}    (\vc ,\vs )
+ \Es   \hNN (\vc ,\vs ) ~, \cr
\partial_t \hw  \,&=\,  \hMM_\w \hw  + \hNN_\w(\vc ,\vs ,\hw ) ~,\cr
}
\EQ{shs}
$$
where, see \equ{13b} and \equ{mmlin2}, with $\hv = \vc+\vs$,
$$
\eqalign{
\hMM \,&=\,\JJ  \MM  \JJ ^{-1}~,\cr
\hat{\cal H} (\vc ,\vs ) \,&=\,\JJ  \MMi  \JJ ^{-1}\hv+ \JJ  \NNi(\JJ ^{-1} \hv)~,\cr
 \hNN (\vc ,\vs ) \,&=\,
 \JJ  \NN(\JJ ^{-1}\hv )~, \cr
\hMM_\w    \,&=\, \widehat  \MM_{\beta ,\tc t }
+ \widehat \MM_{{\bf i},\beta ,\tc t } ~, \cr
\hNN_\w (\vc ,\vs ,\hw ) \,&=\, - 3 \Ustar \cdot  \hv  \starb \hw
-3 \hKc \starb \hv  \starb \hw
 - \hv\starb \hv \starb \hw~.
\cr
}
$$
It is useful to modify this system by introducing the coordinates
$(\uc,\us)$ by
$$\uc\,=\,\vc  , \quad  \us \,=\, - \hMM^{-1}
 (3 \Ustar  \cdot  \vc  \starb \vc ) + \vs ~.
\EQ{coo}
$$
This coordinate transform takes care of the fact that asymptotically
$ \hv_s $ can be expressed by $ \hv_c $. Under the scaling used below
the new variable $ \hu_s $ converges to zero, while
the old variable $ \hv_s $ converges to a nontrivial
expression.

Under this transform \equ{shs} becomes
$$\eqalign{
\partial_t \uc  \,&=\, \hMM  \uc  + \hNN_{\c,{\bf i}}(\uc ,\us )    +\hNN_\c(\uc ,\us ) ,\cr
\partial_t \us  \,&=\, \hMM  \us  +   \hNN_{\s,{\bf i}}(\uc ,\us )
+ \hNN_\s(\uc ,\us )  ,\cr
\partial_t \hw  \,&=\,  \hMM_{\w} \hw   + \hNN_{\w}(\vc ,\vs ,\hw )~,
}
\EQ{shs1}
$$
where
$$
\eqalign{
\hNN_{\c,{\bf i}}(\uc ,\us )  \,&=\, \Ec  \hat{\cal H}
(\uc ,\hMM^{-1}
  \Es   (3 \Ustar  \cdot  \uc  \starb \uc ) + \us )~ ,\cr
\hNN_{\s,{\bf i}}(\uc ,\us )  \,&=\, \Es  \hat{\cal H}
(\uc ,\hMM^{-1}
  \Es   (3 \Ustar  \cdot  \uc  \starb \uc ) + \us )~ ,\cr
 \hNN_\c(\uc ,\us ) \,&=\, \Ec   \hNN
 (\uc ,\hMM^{-1}
  \Es   (3 \Ustar  \cdot  \uc  \starb \uc ) + \us )~ ,\cr
\hNN_\s(\uc ,\us ) \,&=\,  \Es   \hNN
 (\uc ,\hMM^{-1}  \Es
(3 \Ustar   \cdot   \uc  \starb \uc ) + \us ) -
 \partial_t [\hMM^{-1}  \Es
  (3 \Ustar   \cdot    \uc \starb   \uc )] ~.\cr
}
$$
We follow the lines of \sec{Renormalization} and start
with the renormalization process by
introducing the scalings
$$
\eqalign{
\vcn(\kappa,x,\tau ) \,&=\, \uc (\sigma^n\kappa,x,\sigma^{-2n}\tau )~, \cr
\vsn(\kappa,x,\tau )\,&=\, \sigma^{-3n/2} \us
( \sigma^n\kappa,x,\sigma^{-2n}\tau)~, \cr
\hw_{n }(\kappa,x,\tau ) \,&=\, e^{-\gamma \sigma^{-2n} \tau} \hw(
\sigma^n\kappa,x,\sigma^{-2n} \tau)~.\cr
}
$$
(The $3^{\rm rd}$ argument is the time, and the function $w$ has here
another meaning than in \sec{Renormalization}.)
Note again that only the Bloch variable is rescaled, but $x$ is left
untouched.

Under these  scalings
the functions $ \vsn $ and $ w_n $ still converge towards $ 0 $ as
$n\to\infty $.
The variation of constant formula yields now
$$
\eqalignno{
\vcn(\kappa,x,\tau  ) \,&=\, e^{\sigma^{-2n}  \hMM_{\c,n}(\tau  -\sigma^2)}
\hat v_{\c,n-1}(\sigma \kappa ,x,1) \cr
&+
\sigma^{-2n} \int_{\sigma^2}^\tau \d\tau' \,
e^{\sigma^{-2n}  \hMM_{\c,n}(\tau  -\tau')}
\bigl( \hNN_{\c,{\bf i},n}(\vcn,\vsn)  \bigr) (\kappa,x,\tau')
\cr &+
\sigma^{-2n} \int_{\sigma^2}^\tau \d\tau' \,
e^{\sigma^{-2n}  \hMM_{\c,n}(\tau  -\tau')}
\bigl( \hNN_{\c,n}(\vcn,\vsn)\bigr) (\kappa,x,\tau')  ~, \NR{lcl1}
\vsn (\kappa,x,\tau  ) \,&=\, e^{\sigma^{-2n}
\hMM_{\s,n}(\tau  -\sigma^2)} \sigma^{-3/2}
\hat v_{\s,n-1}(\sigma \kappa,x,1) \cr
&+\sigma^{-7n/2} \int_{\sigma^2}^\tau\d\tau'\,
e^{\sigma^{-2n}  \hMM_{\s,n}(\tau  -\tau')}
\bigl( \hNN_{\s,{\bf i},n} (\vcn,\vsn))\bigr) (\kappa,x,\tau') \cr
&+\sigma^{-7n/2} \int_{\sigma^2}^\tau\d\tau'\,
e^{\sigma^{-2n}  \hMM_{\s,n}(\tau  -\tau')}
\bigl( \hNN_{\s,n}(\vcn,\vsn)\bigr) (\kappa,x,\tau') ~, \NR{lcl2}
\hw_n(\kappa,x,\tau) \,&=\,  \hS_{n}(\tau  ,\sigma^2) \hw_{n-1}
(\sigma \kappa,x,1)
\cr
&+ \int_{\sigma^2}^\tau \d\tau'\, \hS_{n}(t,\tau')
\bigl(\hNN_{\w,n}(\vcn,\vsn,\hw_n)\bigr)(\kappa,x,\tau')~,\NR{lcl3}
}
$$
with
$$
\eqalign{
 \hMM_{\c,n} \,&=\, \hLL^n \Ech \hMM  \hLL^{-n} ~, \cr
 \hMM_{\s,n} \,&=\, \hLL^n \Esh \hMM  \hLL^{-n}~ , \cr
\hNN_{\c,{\bf i},n}(\vcn,\vsn ) \,&=\,
\hLL^n
 \hNN_{\c,{\bf i}}(\hLL^{-n}\vcn,\sigma^{3n/2} \hLL^{-n} \vsn ) ~,\cr
\hNN_{\s,{\bf i},n}(\vcn,\vsn ) \,&=\,
\hLL^n
\hNN_{\s,{\bf i}}(\hLL^{-n}\vcn,\sigma^{3n/2} \hLL^{-n}\vsn ) ~,\cr
\hNN_{\c,n}(\vcn,\vsn ) \,&=\, \hLL^n
 \hNN_{\c}(\hLL^{-n}\vcn,\sigma^{3n/2}\hLL^{-n} \vsn ) ~,\cr
\hNN_{\s,n}(\vcn,\vsn ) \,&=\, \hLL^n
\hNN_{\s}(\hLL^{-n} \vcn,\sigma^{3n/2} \hLL^{-n}\vsn )~, \cr
\hNN_{\w,n}(\vcn,\vsn ,\hw_n)
\,&=\, \hLL^n \hNN_{\w}(\hLL^{-n}\vcn,\sigma^{3n/2}\hLL^{-n} \vsn ,
\hLL^{-n}\hw_n)~,\cr
}
$$
where
we recall the definition
$$
\bigl(\hLL \hat f\bigr)(\ell,x)\,\equiv\,\hat f(\sigma
\ell,x)~,
$$
and where
$ \hS_{n}(\tau  ,\tau') $ is now the evolution operator
associated with the equation
$$
\partial_\tau \hat f_{\tau} = \sigma^{-2n} ( \hLL^n  \hMM_{\w}  \hLL^{-n}
+ \gamma) \hat f_{\tau} ~.
\EQ{generatorsh}
$$
Again, the exponential scaling of $ \hw_n $
with respect to time
does not affect
the definition of $ \hNN_{\w} $ due to the fact that $ \hw_n $
only appears linearly.

All this is quite analogous to the developments in Eqs.\equ{new1} and
\equ{new3}.

\SUBSECT{Scaledsh}{The scaled linear evolution operators}First we
bound the linear evolution operators generated by
$ \hMM_{\c,n} $ and $ \hMM_{\s,n} $.

\CLAIM{Lemma}{Mcsnsh}For all $ \rho \in (0,1] $ there exist
$C_{\rho}>0$ and $ \gamma_- > 0 $ such that for $1\ge\tau > \tau '\ge \sigma^2$
and all $ \sigma \in (0,1) $ one has
$$
\eqalignno{
\|e^{\sigma^{-2n}\hMM_{\c,n}(\tau -\tau')} \hLL^n \Ech \hLL^{-n}
\hat g\|_{\KK_{\sigma^n}}\,\le\, &
C(\tau -\tau')^{\rho-1}\|\hat g\|_{\KK_{\sigma^n,\rho}}~, \cr
\|e^{\sigma^{-2n}\hMM_{\s,n}(\tau -\tau')} \hLL^n \Esh \hLL^{-n}
\hat g\|_{\KK_{\sigma^n}}\,\le\, &
C e^{-\gamma_- \sigma^{-2n} (\tau-\tau')}
(\tau -\tau')^{\rho-1}\|\hat g\|_{\KK_{\sigma^n,\rho}}~,
}
$$
for all $ n \in\natural  $.

\PROOF
The first estimate follows directly from the fact that
$$ \hMM_{\c,n}(\ell) f = \mu_1(\ell) \Pc(\ell) f
= - c_1 \ell^2 \Pc(\ell) f + \OO(\ell^3)~.
$$
The second estimate follows from the fact that the
real part of the spectrum
of $ \hMM_{\s,n}(\ell) $ as a function of $ \ell $
can be bounded from above by a strictly negative parabola.
\QED

Next, we bound
$\hS_{n}(\tau ,\tau ')$ as defined through
\equ{generatorsh} and state  the analog of \clm{Snn}.
\CLAIM{Lemma}{Snnsh}Suppose that the stability condition \equ{sc} is satisfied.
Then there is a $ \delta \in (0,1] $ such that
for all $\epsilon '\in(0,1)$ there exists a
$C_{\epsilon '}>0$ such that for $1>\tau > \tau '\ge 0$
and all $ \sigma \in (0,1] $ one has
$$
\eqalign{
\|\hS_{n}(\tau ,\tau ')\hw\|_{\KK_{\sigma^{n}}}\,\le\,  C_{\epsilon'}
\sigma^{-\epsilon' n}e^{-\gamma
\sigma^{-2n}(\tau -\tau ')/ 2} (\tau -\tau ')^{\epsilon'-1}
\|\hw\|_{\KK_{\sigma^{n},\epsilon'}}~,
\cr 
}
\EQ{sboundsh}
$$
for all $ n \in\natural  $.

The proof of \clm{Snnsh}
follows closely the one of \clm{Snn} in \sec{Scaled}.
Therefore, it will be omitted here.
We only remark that the estimate for the solution 
of \equ{generatorsh}
$$
\|\hat f_{\tau} \|_{\hat H^{2,\delta}_0}  \leq C e^{\gamma \sigma^{-2n} 
(\tau -\tau') /2}  
\|\hat f_{\tau'} \|_{\hat H^{2,\delta}_0} 
$$ 
associated to 
\equ{first} can be obtained exactly in the same way as 
\equ{p3bound}.
The estimates for the weights in $ \ell $  and the derivatives with respect to 
$ x $ follow again as in the proof of \clm{Snn}.

\SUBSECT{nonltermsh}{The scaled non-linear terms}Next we estimate the scaled non-linear terms in
$ \NN_{\c,n} $, $ \NN_{\s,n} $, and $ \NN_{\w,n} $.

\CLAIM{Lemma}{nonlemsh}Suppose $ \max\{\| \vcn \|_{\KK_{\sigma^n}} , \| \vsn \|_{\KK_{\sigma^n}},
 \| \hw_n \|_{\KK_{\sigma^n}} \}\leq 1 $. Then
 for all $\epsilon '\in(0,1)$ there exist
$C_1,C_{\epsilon '}>0$
such that for
all $ \sigma \in (0,1] $ one has
$$
\eqalignno{
\|\hNN_{\c,n} \|_{\KK_{\sigma^n,1/4}}\,\le\, & C_1
\sigma^{5n/2} (\| \vcn \|_{\KK_{\sigma^n}}  +
\| \vsn \|_{\KK_{\sigma^n}} )^2
\cr
\|\hNN_{\s,n} \|_{\KK_{\sigma^n,1/2}}\,\le\, & C_1
\sigma^{2n} (\| \vcn \|_{\KK_{\sigma^n}}  +
\| \vsn \|_{\KK_{\sigma^n}} )^2
\cr
\|\hNN_{\w,n} \|_{\KK_{\sigma^n,\epsilon'}}\,
\,\le\, & C_{\epsilon '} \sigma^{(1-\epsilon') n} (\| \vcn \|_{\KK_{\sigma^n}}  +
\| \vsn \|_{\KK_{\sigma^n}} )  \| \hw_n \|_{\KK_{\sigma^n}} ~.
}
$$

\PROOF Throughout the proof we use
$$
\eqalign{
\bigl(\widehat \LL (\hat f\starb \hat g)\bigr)(\kappa)\,
=\,\sigma \bigl((\widehat \LL \hat f)\starb
(\widehat \LL \hat g)\bigr)(\kappa)~.
}\EQ{id1sh}
$$

i) We start with the estimates for $ \hNN_{\w,n} $.
The most dangerous term in
$$
\hNN_\w (\vc ,\vs ,\hw ) \,=\,  3 \Ustar \cdot  \hv  \starb \hw
-3 \hKc \starb \hv  \starb \hw
 - \hv\starb \hv \starb \hw
$$
is $ 3 \hKc \starb \hv  \starb \hw $. From \equ{id1sh} we obtain
a $ \sigma^n $ for the scaled version of $ \hv  \starb \hw $.
We loose $ \sigma^{-\epsilon' n} $ by  taking the norm
in $ \KK_{\sigma^n,\epsilon'} $
due to the fact that
$ \hKc $ is fixed and does not scale when time evolves.

ii) We use again \equ{id1sh} to obtain the estimates for $\hNN_{\s,n} $.
The only difficulty
stems from the term
$$
\partial_t [\hMM^{-1}  \Es
  (3 \Ustar   \cdot    \uc \starb   \uc )]
  =
\hMM^{-1}  \Es
  (6 \Ustar   \cdot    \uc \starb   \partial_t \uc )
$$
coming from the change of coordinates \equ{coo}.
This can be estimated in the required way by
expressing $ \partial_t \uc $ by the right hand side of
\equ{shs1}, by
using then the points  ii.1)--ii.3) and  the fact we already have
a factor $ \sigma^n $ by $ \uc \starb   \partial_t \uc $ using again
\equ{id1sh}.

ii.1) The first bound for the terms on the right hand side of \equ{shs1}
 is
$$ \| \hMM_{c,n} \vcn\|_{\KK_{\sigma^n,\rho}} \leq
C \sigma^{2(1-\rho)n} \| \vcn \|_{\KK_{\sigma^n}}~,
$$
(with $ \rho=1/2 $ for our purposes)
which follows from the form of $ \mu_1(\ell) $
by using the following lemma.
\CLAIM{Lemma}{derivative}Let
$ \mu \in \CC^2_{\rm per}([-1/2,1/2),\CC^2((0,2\pi),\complex)) $ with
$ \| \mu(\ell,\cdot) \|_{\CC^2((0,2\pi),\complex)}
\leq C |\ell|^{2(1-\rho)} $ for a $ \rho \in [0,1] $. Then, there exists
a $ C > 0 $ such that
for all $ \sigma \in (0,1] $ we have
$$ \| (\hLL_{\sigma} \mu  ) \hu \|_{\KK_{\sigma,\rho}}
\leq C \sigma^{2(1-\rho)}
\| \mu \|_{\CC^2_{\rm per}([-1/2,1/2),\CC^2((0,2\pi),\complex))}
\| \hu \|_{\KK_{\sigma}}~.
\EQ{estans}
$$

\PROOF This follows since
$$ \sup_{\ell \in \real} |
{{\ell^{2(1-\rho)} \sigma^{2(1-\rho)}
}\over{
(1+\ell^2)^{(1-\rho)}}} | < C \sigma^{2(1-\rho)}~.
$$
\QED

ii.2)
By \clm{linintlemsh} below the term $ \hNN_{\c,{\bf i},n} $
is exponentially small in terms of $ \sigma $.

ii.3)
From \equ{id1sh} we easily obtain
$$
\| \hNN_{\c,n} \|_{\KK_{\sigma^n}} \leq
\sigma^{n} (\| \vcn \|_{\KK_{\sigma^n}}  +
\| \vsn \|_{\KK_{\sigma^n}} )^2~.
$$

iii)
From [Schn96] we recall
the estimates for
the $ \hNN_{\c,n} $ part.
Note that $\hNN_{\c,n}$ can be written as
$$
\hNN_{\c,n} \,=\,
\hat s_1 + \hat s_2 + \hNN_{\c,n,{\rm r}}~,
$$
where
$$
\eqalign{
\hat s_1 & =  - 3  \sigma^{n} \hLL^n \Ec
 \hLL^{-n} (\Ustar  \cdot \vcn  \starb
\vcn )~, \cr
\hat s_2 & =  - 6  \sigma^{2n}  \hLL^n \Ec
\hLL^{-n}(\Ustar  \cdot   \vcn
 \starb
(\widehat\MM_{\s,n})^{-1}
(3 {\Ustar } \cdot  \vcn  \starb \vcn )) \cr & - \sigma^{2n}
\hLL^n \Ec
\hLL^{-n} (\vcn  \starb \vcn  \starb \vcn )~, \cr
\| \hNN_{\c,n,{\rm r}}  \|_{\KK_{\sigma^n}} & =  \OO(\sigma^{5n/2}
(\| \vcn  \|_{\KK_{\sigma_n}}
+ \| \vsn  \|_{\KK_{\sigma^n}})^2)~.
}
$$
The estimate for $  \hNN_{\c,n,{\rm r}} $
follows easily by applying again \equ{id1sh}.

It remains to estimate $ \hat s_1 $ and $ \hat s_2 $.
These estimates have been obtained in [Schn96]. For completeness we
recall some of the arguments.
Introducing $ a_n(\ell) \in \complex  $ by $ \vcn(\ell,x) = a_n(\ell)
 \varphi_{\sigma^n \ell}(x) $ shows that
the terms $ \hat s_1 $ and $ \hat s_2 $ are of the form
$$
\eqalign{
\hat s_2(\ell,x) \,&=\, \bigg(\sigma^{2n}
\int \dm \int \dk K_2(\sigma^{n}\ell,\sigma^{n}(\ell-m),
 \sigma^{n}(m-k),\sigma^{n}
k) \cr  &   \times
a_n(\ell-m) a_n(m-k) a_n(k)  \biggr)\,
\varphi_{\sigma^n \ell}(x) ~,\cr
\hat s_1(\ell,x)\,&=\, \biggl(\sigma^{n} \int \dm K_1( \sigma^{n}\ell,
 \sigma^{n}(\ell-m),\sigma^{n}m)\, a_n(\ell-m)\, a_n(m) \biggr)
\varphi_{\sigma^n \ell}(x)~,\cr
}
$$
with $ K_j : \real^{2+j} \rightarrow \complex $ the kernel of an
integral operator.
The detailed expression for $ K_1 $ is given in \equ{K1ans} below.

The case $ n = m = k = \ell = 0  $ corresponds to the spatially periodic case.
In the spatially periodic case there exists a center manifold
$$
\Gamma = \{u = U_{0,a} \ | \ a \in \real \}~,
$$
consisting of the spatially periodic fixed points
related to each other by the translation invariance
of the original Swift-Hohenberg equation.
By a formal calculation it turns out that the flow
of the one-dimensional center manifold $ \Gamma $ is determined
by the ordinary differential equation
$$
{{d}\over{dt}} {a} = 0 \cdot a + K_1(0,0,0) a^2 +   K_2(0,0,0,0) a^3
+ \OO(a^4)~.
$$
Since the center manifold consists of fixed points the flow $ a = a(t) $
is trivial, \ie $ {{d}\over{dt}} {a} = 0 $. Consequently, we obtain
$ K_1(0,0,0) = K_2(0,0,0,0) = 0 $. Therefore,
$$
 |K_2(\ell,\ell-m,m-k,k)| \,\leq\, C (|\ell|+|\ell-m|+|m-k|+|k|)~,
$$
and so \equ{id1sh} and \equ{estans} imply
$$
\| \hat s_2 \|_{\KK_{\sigma^n,1/2}} \,\leq\,
C \sigma^{3n}
(
\| \vcn   \|_{\KK_{\sigma^n}}
+ \| \vsn  \|_{\KK_{\sigma^n}} )^2 ~.
$$
Interestingly it turned out that
the first derivatives  of $ K_1 $ vanish as well.
Since the eigenvalue problem $ \MM_{\ell} \varphi_ {\ell}= \mu_1 ({\ell})
\varphi_ {\ell} $ is self-adjoint,
the projection $ \Pc (\ell) $ is orthogonal in $ \L^2(0,2\pi) $ and is
given by
$ \Pc (\ell) u = (\int \overline{\varphi_ {\ell}(x)}
u(\ell,x) dx) \varphi_ {\ell}( \cdot) $. Thus
$$ K_1(\ell,\ell-m,m) = 3
\int \dx \overline{\varphi_ {\ell}(x)} \varphi_ {\ell-m}(x) \varphi_ {m}(x)
 U(x)~.
\EQ{K1ans}
$$
Expanding $
\varphi_{\ell}(x)  = \partial_x U(x) + i \ell g(x) + \OO(\ell^2),
$
with $ g(x) \in \real $
yields
$$
\eqalign{
 K_1(\ell,\ell-m,m) = & 3 \int \dx \bigg((\partial_x U(x))^3
U(x)
\cr
 & - i\ell g(x)(\partial_x U(x))^2
 U(x) +
 i(\ell-m) g(x)(\partial_x U(x))^2
 U(x)  \cr & +
(\partial_x U(x))^2 i m g(x)
 U(x)
 + \OO( \ell^2 + (\ell-m)^2 + m^2 )\bigg)~.
}
$$
Note that $
U(x) $ is an even function,
so $ \partial_x U $ is odd, which proves again $ K_1(0,0,0) = 0 $.
Since, in addition,
the first order terms cancel we have
$$
| K_1(\ell,\ell-m,m)| \,\leq\, C
| \ell^2 + (\ell-m)^2 + m^2 |~,
$$
and so from \equ{id1sh} and \equ{estans}
$$ \| \hat s_1 \|_{\KK_{\sigma^n,1/4}} \,\leq\,
C \sigma^{5n/2}
( \| \vcn\|_{\KK_{\sigma^n}}
+ \| \vsn  \|_{\KK_{\sigma^n}} )^2~. $$
Summing the estimates shows the assertion.
\QED

\SUBSECT{integral}{Bounds on the integrals}Here we estimate the
integrals in the variation of constant formula
in terms of the
following quantities.
\CLAIM{Definition}{Rnfull}For all $n$, we define
$$
R_{\c\s,n}^u\,=\,\sup_{\tau \in[\sigma^2,1]}\|
\hv_{\c,n}(\tau )\|_{\KK_{\sigma^n}}
+\sup_{\tau \in[\sigma^2,1]}\|\hv_{\s,n}(\tau
)\|_{\KK_{\sigma^n}}~,
\quad\text{and}\quad
R_n^w\,=\,\sup_{\tau \in[\sigma^2,1]}\|\hw_n(\tau )\|_{\KK_{\sigma^n}}~.
$$

In the following two lemmas
we estimate the integrals appearing
in \equ{lcl1}--\equ{lcl3}.
\CLAIM{Lemma}{nonintlemsh}Assume $ R_{\c\s,n}^u + R_n^w \leq 1 $. Then
for all  $1\geq \tau  \ge \sigma^2$
and all $ \sigma \in (0,1] $ one has
$$
\eqalignno{
\| \sigma^{-2n} \int_{\sigma^2}^\tau \d\tau' \,
e^{\sigma^{-2n}  \hMM_{\c,n}(\tau  -\tau')}
\bigl( \hNN_{\c,n}(\vc,\vs)\bigr) (\cdot,\cdot,\tau') \|_{\KK_{\sigma^n}} & \leq
C \sigma^{n/2 } (R_{\c\s,n}^u)^2   ~,
\cr
\|\sigma^{-7n/2} \int_{\sigma^2}^\tau\d\tau'\,
e^{\sigma^{-2n}  \hMM_{\s,n}(\tau  -\tau')}
\bigl( \hNN_{\s,n}(\vc,\vs)\bigr) (\cdot,\cdot,\tau') \|_{\KK_{\sigma^n}}& \leq
 C \sigma^{n/2 } (R_{\c\s,n}^u)^2~,
\cr
\| \int_{\sigma^2}^\tau \d\tau'\, \hS_{n}(t,\tau')
\bigl(\hNN_{\w,n}(\vc,\vs,\hw)\bigr)(\cdot,\cdot,\tau')\|_{\KK_{\sigma^n}}& \leq~
C \sigma^{n(1-\epsilon')} R_{\c\s,n}^u R_n^w~. }
$$

\PROOF We first use \clm{Mcsnsh} and \clm{nonlemsh}.
For the second integral in \equ{lcl1} we get a bound
$$
\eqalign{
&
\sup_{\tau \in [\sigma^2,1]}
\| \sigma^{-2n} \int_{\sigma^2}^\tau
\d \tau' e^{\sigma^{-2n}   \widehat\MM_{\c,n} (\tau-\tau')} \bigl(
\hNN_{\c,n}(\vc,\vs)\bigr) (\cdot,\cdot,\tau')  \|_{
\KK_{\sigma^n}}
\cr
& \,\leq\, C \sigma^{-2n}
(R_{\c\s,n}^u)^2  \sigma^{5n/2}
\int_{\sigma^2}^1 \d \tau' (1-\tau')^{-3/4}   \cr
& \,\leq\, C \sigma^{n/2 } (R_{\c\s,n}^u)^2~.}
$$
For the second integral in \equ{lcl2} we find similarly
$$
\eqalign{
\sup_{\tau \in[\sigma^2,1]}&
\|\sigma^{-7n/2}\int_{\sigma^2}^\tau \d\tau '\,
e^{\sigma^{-2n}\hMM_{\s,n}(\tau -\tau')}
\bigl( \hNN_{\s,n}(\vc,\vs)\bigr) (\cdot,\cdot,\tau')
\|_{\KK_{\sigma^n}}\cr
\,&\le\,C (R_{\c\s,n}^u)^2\sigma^{-3n/2}
\int_{\sigma^2}^1 \d\tau '\,
e^{-C\sigma^{-2n}(1-\tau ')} (1-\tau')^{-1/2} \cr
\,&\le\,
C \sigma^{n/2} (R_{\c\s,n}^u)^2~.\cr
}
$$
For the  integral in \equ{lcl3} we find, using now \clm{Snnsh}
and \clm{nonlemsh},
a bound
$$
\eqalign{
C \sigma^{-2n} \int _{\sigma^2 }^\tau& \d \tau '\,
(\sigma^{- \epsilon' n/2}e^{-\gamma \sigma^{-2n}
(\tau -\tau ' )/2
 } (\tau -\tau ' )^{\epsilon'/2-1})
(\sigma^{(1- \epsilon' /2)n} R_{\c\s,n}^u R_n^w) \cr
\,&\le\,
C \sigma^{n(-1-\epsilon')}\sigma^{2n} R_{\c\s,n}^u R_n^w\,\le\,
C \sigma^{n(1-\epsilon')} R_{\c\s,n}^u R_n^w
~.
}
$$
\QED

\CLAIM{Lemma}{linintlemsh}Assume $ R_{\c\s,n}^u + R_n^w \leq 1 $. Then
for all  $1 \geq \tau  \ge \sigma^2$
and all $ \sigma \in (0,1) $ one has
$$
\eqalignno{
\|\sigma^{-2n} \int_{\sigma^2}^\tau \d\tau' \,
e^{\sigma^{-2n}  \hMM_{\c,n}(\tau  -\tau')}
\bigl( \hNN_{\c,{\bf i},n} (\vc,\vs) \bigr) (\cdot,\cdot,\tau')\|_{\KK_{\sigma^n}}
& \leq\, \,C e^{-(\beta
(c-\tc)+\gamma) \sigma^{-n}} R_n^w~, \cr
\|\sigma^{-7n/2} \int_{\sigma^2}^\tau\d\tau'\,
e^{\sigma^{-2n}  \hMM_{\s,n}(\tau  -\tau')}
\bigl( \hNN_{\s,{\bf i},n}(\vc,\vs) \bigr) (\cdot,\cdot,\tau')\|_{\KK_{\sigma^n}}
& \leq\, \,C e^{-(\beta
(c-\tc)+\gamma) \sigma^{-n}} R_n^w~.
}
$$

\PROOF We restrict ourselves to  the linear part $ \MMi $.
A typical term  of \equ{lcl1}---the first in the definition of $\MMi$ in \equ{13b}---can be rewritten as
$$
\eqalign{
\sigma^{-2n}&\biggl(\int_{\sigma^2 }^{\tau } \d \tau '\,
e^{\widehat\MM_{\c,n} (\tau -\tau ')}
\hLL^n\bigl (\hK_{c\sigma^{-2n} \tau '}\starb (\hLL^{-n} \hat u_{n,\tau '})\bigr
)\biggr)(\kappa,x)\,U(x) \cr
\,&=\,
\sigma^{-2n}\biggl(\int_{\sigma^2 }^{\tau }\d \tau '\, e^{\widehat\MM_{\c,n} (\tau -\tau ')}
\hLL^n\bigl (\hK _{c\sigma^{-2n} \tau '}\starb \hat u_{\sigma^{-2n}\tau '}\bigr
)\biggr)(\kappa,x)\,U(x)
~. \cr
}
$$
Note next that
$$
\eqalign{
\bigl(\hK _{c\sigma^{-2n} \tau '}&\starb \hat u_{\sigma^{-2n}\tau '}\bigr
)(\kappa,x)\cr\,&=\,
\int \d\ell\,
\hK _{c\sigma^{-2n} \tau '}(\kappa -\ell-i\beta,x)\, \hw(\ell,x,\sigma^{-2n}\tau
')e^{-i\ell\tc \sigma^{-2n}\tau'} e^{-\gamma \sigma^{-2n} \tau'} \cr
& \times e^{-\beta (c - \tc)  \sigma^{-2n} \tau'}
e^{i(\kappa -\ell) c \sigma^{-2n} \tau'}~.
\cr
}
$$
Using this identity, we get (because
$\exp(\widehat\MM_{\c,n}(\tau-\tau'))$ is bounded):
$$
\eqalign{
\sigma^{-2n}&\,\|\int_{\sigma^2}^\tau \d \tau'\,
e^{\widehat\MM_{\c,n}(\tau-\tau')}
\hLL^n\bigl(\hK _{c\sigma^{-2n}\tau '}\starb
\hu_{n,\tau' }\bigr)\|_{\KK_{\sigma^n}}\cr
\,&\le\,C\sigma^{-2n} \int_{\sigma^2  }^{\tau } \d\tau '
\|\hLL^n(\hK_{c\sigma^{-2n}\tau '}\starb
\hu_{n,\tau' })\|_{\KK_{\sigma^n}}\cr
\,&\le\,C\sigma^{-2n} \int_{\sigma^2  }^{\tau } \d\tau 'e^{-\beta
(c-\tc)\sigma^{-2n}\tau '}
\|(\kappa,x)\mapsto e^{-ic\sigma^{-2n}\tau '\kappa}\hK_{c\sigma^{-2n}\tau '}(
\kappa- i\beta,x)\|_{\KK_{\sigma^n}} \cr
&~~~~~~~~~~~~~~~~~~~~~\times\,\|(\kappa,x)\mapsto e^{-i\kappa\tc
\sigma^{-2n}\tau'}\hw_{n,\tau' }(\kappa,x)\|_{\KK_{\sigma^n}}
 e^{-\gamma \sigma^{-2n} \tau'}\cr
\,&\le\,C\sigma^{-2n} \int_{\sigma^2 }^{\tau } \d\tau '
(1+\tc\sigma^{-2n} \tau ')^2(1+c\sigma^{-2n} \tau ')^2  e^{-\beta
(c-\tc)\sigma^{-2n}{\tau'}} e^{-\gamma \sigma^{-2n} \tau'} R_n^w
\cr
\,&\le\,C \sigma^{-6n}e^{-(\beta
(c-\tc)+\gamma)\sigma^{-2(n-1)}} R_n^w\,\,\leq\, \,C e^{-(\beta
(c-\tc)+\gamma) \sigma^{-n}} R_n^w~. \cr
}\EQ{bound555sh}
$$
The non-linear terms coming from $ \NNi $ can be handled in exactly
the same way and yield similar bounds. The same is true for the terms with
$\NN_{\s,{\bf i},n} $ in \equ{lcl2}.
\QED

\SUBSECT{scaleinisec}{Bounds on the initial condition}Here, we estimate
the first terms
on the right hand side of
the variation of constant formulae \equ{lcl1}--\equ{lcl3}.

\CLAIM{Lemma}{scaleini}For all  $1 \geq \tau  \ge \sigma^2$
and all $ \sigma \in (0,1] $
we have
$$
\eqalign{
\|e^{\sigma^{-2n}\hMM_{\c,n}(\tau -\sigma^2)}\hLL^n \Ech
\hLL^{-n} \hLL \hat g\|_{\KK_{\sigma^n}}
& \,\le\,
C\sigma^{-5/2} \|\hat g\|_{\KK_{\sigma^{n-1}}}~,
\cr
\|e^{\sigma^{-2n}\hMM_{\s,n}(\tau -\sigma^2)}
\hLL^n \Esh \hLL^{-n} \sigma^{-3/2} \hLL \hat g\|_{\KK_{\sigma^n}}
& \,\le\,
C \sigma^{-4}
e^{-C\sigma^{-2n}(\tau-\sigma^2) } \|\hat g\|_{\KK_{\sigma^{n-1}}}~,\cr
\|\hS_{n}(\tau,\sigma^2 )
\hLL \hat g
\|_{\KK_{\sigma^n}}
&\,\le\, C\sigma^{-5/2}\sigma^{-\epsilon' n}e^{-\gamma
\sigma^{-2n}(\tau-\sigma^2 )/
2}\|\hat g \|_{\KK_{\sigma^{n-1}}}~.\cr
}
$$

\PROOF As before we have
$$
\|\widehat \LL \hat f\|_{\KK_{\sigma^n}}
\,\le\, \sigma^{-5/2} \|\hat f\|_{\KK_{\sigma^{n-1}}}~,
\EQ{nuern1}
$$
for $0<\sigma\leq 1$.
Therefore, the first two bounds of \clm{scaleini} follow immediately
from \clm{Mcsnsh}.
The third inequality is a little less obvious:
First note that
$$ \hS_{n}(\tau,\sigma^2 )
\hw _{n}(\cdot,\cdot,\sigma^2)=
\hat\LL\bigl(\hS_{n-1}(\tau \sigma^{-2},1)
\hw _{n-1}(\cdot,\cdot,1) \bigr)~.$$
Therefore,
$$
\eqalign{
\,& \,\|\tilde\LL\bigl(\hS_{n-1}(\tau \sigma^{-2},1)
\hw _{n-1}(\cdot,\cdot,1)\bigr)\|_{\KK_{\sigma^n}} \cr
\,&\,\leq\, \,\sigma^{-5/2}\|
\hS_{n-1}(\tau \sigma^{-2},1)
\hw _{n-1}(\cdot,\cdot,1)
\|_{\KK_{\sigma^{n-1}}} \cr
\,&\le\, C\sigma^{-5/2}\sigma^{-\epsilon' n}e^{-\gamma
\sigma^{-2n}(\tau-\sigma^2 )/
2}\|\hw _{n-1}(\cdot,\cdot,1)\|_{\KK_{\sigma^{n-1}}}~.\cr
}
\EQ{sboundnminus1sh}
$$
The claim is now an immediate consequence of \clm{Snnsh}.
\QED

\SUBSECT{apriorish}{A priori bounds on the non-linear problem}This
section follows closely \sec{apriori}.
We need a priori bounds on the
solution of \equ{lcl1}--\equ{lcl3}. We (re)define now quantities
analogous to those of \clm{Rn}.
\CLAIM{Definition}{Rhonfull}For all $n \in \natural $, we define
$$
\rho _{\c\s,n}^u\,=\,\| \hv_{\c,n}|_{\tau=1
}\|_{\KK_{\sigma^n}}+\|\hv_{\s,n}|_{\tau=1 } \|_{\KK_{\sigma^n}}~,\quad\text{and}\quad
\rho_{n}^w\,=\,\|\hw_n|_{\tau=1 }\|_{\KK_{\sigma^n}}~.
$$

\CLAIM{Lemma}{apriorish}For all
$n \in \natural $ there is a constant $\eta_n>0$
such that  the following holds:
If $\rho _{\c\s,n-1}^u$, $\rho_{n-1}^w$, and
$\sigma>0$ are smaller than $\eta_n$, the solutions of \equ{lcl1}--\equ{lcl3}
exist for all $\tau \in[\sigma^2,1]$. Moreover, we have the estimates
$$
R_{\c\s,n}^u\,\le\,
C \sigma^{-4} \rho _{\c\s,n-1}^u + C e^{-C \sigma^{-n}}
 R_n^w +
C \sigma^{n/2} (R_{\c\s,n}^u)^2  ~,
\EQ{after1sh}
$$
and
$$
R_n^w\,\le\,
C \sigma^{-5/2- \epsilon' n}{\rho}_{n-1}^w +
C \sigma^{n(1-\epsilon')} R_{\c\s,n}^u R_n^w ~,
\EQ{after2sh}
$$
with a constant $C$ independent of $\sigma$ and $n$.

\REMARK We remark again that there is no need for a detailed expression
for $ \eta_n $ since the existence of
the solutions is guaranteed if we can show
$ R_{\c\s,n}^u< \infty $ and $ R^w_n < \infty $.
By \equ{after1sh} and \equ{after2sh} we have detailed control of these
quantities in terms of the norms of the initial conditions and $ \sigma $.

\PROOF For the derivation of the estimates we assume in the sequel,
without loss of generality, that
$R_{\c\s,n}^u+R_n^w\le 1$.
For the first term in \equ{lcl3} we obtained in \clm{scaleini} a bound
$$
C\sigma^{-5/2}  \sigma^{- \epsilon' n}
\rho_{n-1}^w~.
\EQ{exposh}
$$
For the second term in \equ{lcl3}, we
obtained in \clm{nonintlemsh} a bound
$ C \sigma^{n(1-\epsilon')} R_{\c\s,n}^u R_n^w $.

We now discuss in detail \equ{lcl2}.
Using \clm{scaleini} the first term is bounded by $
C \sigma^{-4} \rho _{\c\s,n-1}^u $.
\clm{nonintlemsh} and \clm{linintlemsh} yield
for the second and third terms a bound
$ C \sigma^{n/2} (R_{\c\s,n}^u)^2 + C e^{-C \sigma^{-n}} R_n^w~ $
for a $ C >0 $ independent of $ \sigma \in (0,1] $ and
$ n \in \natural $.

Finally, we come to the bounds
for \equ{lcl1}.
Using \clm{scaleini} the first term is bounded by $
C \sigma^{-5/2} \rho _{\c\s,n-1}^u $.
\clm{nonintlemsh} and \clm{linintlemsh} yield
for the second and third terms a bound
$ C \sigma^{n/2} (R_{\c\s,n}^u)^2 + C e^{-C \sigma^{-n}} R_n^w~ $
for a $ C >0 $ independent of $ \sigma \in (0,1] $ and
$ n \in \natural $.

The proof of \clm{apriorish} now follows by applying the contraction
mapping principle to the system consisting of
\equ{lcl1}, \equ{lcl2}, and \equ{lcl3}.

Then for $ \rho _{\c\s,n-1}^u $, $ \rho_{n-1}^w $ and $\sigma>0$
 sufficiently
small the Lipschitz constant on the right hand side
of \equ{lcl1} to \equ{lcl3}
in $\CC([\sigma^2,1],\KK_{\sigma^n}) $
is  smaller than 1. An application of a classical fixed point argument
completes the proof of \clm{apriorish}.\QED

\SUBSECT{IP}{The iteration process}As in the case of the simplified
problem,
we decompose the solution $\vcn (\cdot,\cdot,\tau)$ for $ \tau= 1 $
into a Gaussian part and a
remainder. Let $\tpsi (\kappa)\,=\,e^{-c_1 \kappa^2}$ and write
$$
\vcn (\kappa,x,1)\,=\,A_{n } \tpsi (\kappa) \varphi_{\sigma^{-n} \kappa}(x)
+ \hr_{n }(\kappa,x)~,
$$
where
$\hr_{n }(0,x)=0$, and the amplitude $A_{n }$ is in $\complex$.
We also define $\hPi:\KK_{\sigma} \to\complex$ by
$$
(\hPi f) \varphi_0= \Pc(0) f\big|_{\kappa=0}~.
\EQ{pidefsh}
$$
Then \equ{lcl1} can be decomposed accordingly and takes the form
$$
\eqalignno{
A_{n}\,&=\,
A_{n-1} + \hPi\bigg(\int_{\sigma^2
}^{1}\d\tau'\,e^{\sigma^{-2n} \widehat\MM_{\c,n} (1-\tau')}\bigl(
\sigma^{-2n}
(\hNN_{\c,{\bf i},n}+\hNN_{\c,n})\bigr)\biggr)~,\NR{ansh}
\hr
_{n }(\kappa,x)\,&=\,e^{\sigma^{-2n} \widehat\MM_{\c,n}(1-\sigma^2)}
\hr_{n-1}(\sigma\kappa,x)\cr&~~~~~~~~~~~~~~~+
\sigma^{-2n}
\int_{\sigma^2
}^{1}\d\tau'\,\left (e^{\sigma^{-2n} \widehat\MM_{\c,n}(1-\tau')}
(\hNN_{\c,{\bf i},n}+\hNN_{\c,n})
\right)(\kappa,x)\NR{rnsh}
&~~~~~~~~~~~~~~~+e^{\sigma^{-2n} \widehat\MM_{\c,n}(1 -\sigma^2)}
A_{n-1 } \tpsi (\sigma\kappa) \varphi_{\sigma^{-n} \kappa}(x) -
A_{n}\tpsi (\kappa)\varphi_{\sigma^{-n} \kappa}(x) ~.\cr
}
$$
If we define next $ \rr_n = \| \hr_{n } \|_{\KK_{\sigma^n}}
+ \| \vsn |_{\tau=1}  \|_{\KK_{\sigma^n}}  $ then the above
construction  implies
$ \rho_{\c\s,n}^u \leq C (|A_n| + \rr_{n}) $.

Our main
estimate is now
\CLAIM{Proposition}{decaysh}There is a constant $C>0$ such that
for sufficiently small $ \sigma > 0 $
the solution $(v_{\c,n},v_{\s,n},w_n) $ of \equ{lcl1}--\equ{lcl3}
satisfies for all $n \in \natural $:
$$
\eqalignno{
|A_{n}-A_{n-1}|\,&\le\,C
e^{-C \sigma^{-n}} R_{n}^w +C\sigma^{n/2}
(R_{\c\s,n}^u)^2~,\NR{a1sh}
\rr_n
\,&\le\, \rr_{n-1}/2 + C e^{-C \sigma^{-n}} R_{n}^w
+C\sigma^{n/2 }(R_{\c\s,n}^u)^2 + C \sigma^n R_{\c\s,n}^u~, \NR{a2sh}
\rho_n^w
\,&\le\, C e^{- C \sigma^{-2n}} \rho_{n-1}^w +
C\sigma^{n (1-\epsilon')
}R_{\c\s,n}^u R_n^w~.\NR{a3sh}
}
$$

\PROOF We begin by bounding the difference $A_{n}-A_{n-1}$  using \equ{ansh}.
Since $ \hat{f}$ is in $H^2 $ as a function of $ \ell $ we obviously have
$$
|\widehat \Pi\hat  f|\,\le\,C \|\hat f\|_{\KK_{\sigma^n}}~.
\EQ{Piboundsh}
$$
Thus, it suffices
to bound the norm of the integral in \equ{ansh},
but this has already been done in the proof of \clm{nonintlemsh}
and \clm{linintlemsh}.

We next bound $\hat r_{n}$ in terms of $\hat r_{n-1 }$, using
\equ{rnsh}. The
first term is the one where the projection is crucial:
For  $\sigma>0 $ sufficiently small,
$\hat r_{n-1} \in\KK_{\sigma^{n-1}} $ with $\hat r_{n-1}(0)=0$ one has
$$
\| (\kappa,x) \mapsto
e^{\sigma^{-2n} \widehat\MM_{\c,n}(1-\sigma^2)}
\hr_{n-1}(\sigma\kappa,x)
\|_{\KK_{\sigma^n}}
\,\le\,  \HALF\|\hat r_{n-1}\|_{\KK_{\sigma^{n-1}}}~,
\EQ{crucialsh}
$$
as in the proof of \clm{decay}.
This leads for the  first term
in \equ{rnsh} to
a bound (in $\KK_{\sigma^n}$)
$$
 \rho_{n-1}^r/2~.
\EQ{bound1sh}
$$
The second and third term have been bounded in
the proof of \clm{nonintlemsh} and \clm{linintlemsh} by
$$
C e^{-C \sigma^{-n}} \Ru_n +C\sigma
^{n/2}(\Ru_n)^2~.
\EQ{bound67sh}
$$
Finally, the last term
$$
\widehat X_{n}(\kappa,x)\,\equiv\,
e^{\sigma^{-2n} \widehat\MM_{\c,n}(1 -\sigma^2)}
A_{n-1 } \tpsi (\sigma\kappa) \varphi_{\sigma^{-n} \kappa}(x) -
A_{n}\tpsi (\kappa)\varphi_{\sigma^{-n} \kappa}(x)~,
$$
in \equ{rnsh} leads to a bound (in
$\KK_{\sigma^n}$):
$$
\|\widehat X_{n }\|_{\HHvoid}\,\le\,C e^{-C \sigma^{-n}} R_{n-1}^w +C\sigma
^{n/2}(R_{\c\s,n}^u)^2  + C \sigma^n R_{\c\s,n}^u~,
\EQ{b7sh}
$$
where the last term is due to $ \mu_1(\ell) = - c_1 \ell^2 + \OO(\ell^3) $
not being exactly a parabola.
For details see [Schn96].
Collecting the bounds, the assertion \equ{a2sh}
for $\hr_{n }$ follows.
Finally, the bounds on $\rho_{n }^w$ follow the in the same way as those in
\clm{apriorish}.
The proof of \clm{decaysh} is complete.
\QED

\LIKEREMARK{Proof of \clm{main2}}As before the proof is just an induction
argument, using repeatedly
the above estimates.
Again we write $ C $ for constants which can be chosen independent
of $ \sigma $ and $ n $.
Assume that $ R = \sup_{n \in\natural } R_{\c\s,n}^u < \infty $ exists.
{}From \clm{apriorish}
we observe for $ \sigma > 0 $ sufficiently small,
$$
\eqalign{
R_n^w  \,&\,\leq\, \, {{C \sigma^{-5/2 - n \epsilon'} \rho_{n-1}^w}\over
{1- C \sigma^{n(1-\epsilon')} R}}
\,\leq\, C \sigma^{-5/2 - n \epsilon'} \rho_{n-1}^w ~,\cr
R_{\c\s,n}^u  \,&\,\leq\, \,
{{C \sigma^{-4} \rho _{\c\s,n-1}^u + C e^{-C \sigma^{-n}}
R_n^w}\over
         {1- C \sigma^{n/2} R} }
\cr
\,&\,\leq\, \,
C \sigma^{-4}
\rho _{\c\s,n-1}^u + C e^{-C \sigma^{-n}} \rho_{n-1}^w~,
}
\EQ{ppsh}
$$
with a constant $ C $ which can be chosen independent
of $ R $.
Using \clm{decaysh} we find
$$
\eqalign{
|A_{n}-A_{n-1}|\,&\le\,C
e^{-C \sigma^{-n}} \rho_{n-1}^w +C\sigma
^{n/2} \sigma^{-4} \rho_{\c\s,n-1}^u~, \cr
\rr_n
\,&\le\, \rr_{n-1}/2 + C e^{-C \sigma^{-n}} \rho_{n-1}^w +C \sigma
^{n/2}\sigma^{-4} \rho_{\c\s,n-1}^u~, \cr
\rho_{\c\s,n}^u & \,\leq\, C (|A_n| + \rr_n) ~,\cr
\rho_n^w\,&\le\,  C e^{- C \sigma^{-2n}} \rho_{n-1}^w +
C\sigma^{n (1-\epsilon')} \sigma^{-5/2 - n \epsilon'} \rho_{n-1}^w~.
}
$$
Therefore, we can choose $ \sigma > 0 $ so small that
for $ n > 9 $:
$$
\eqalignno{
|A_{n}-A_{n-1}| \,&\, \leq
\rho_{n-1}^w/10 +\sigma^{n-9} (|A_{n-1}| + \rr_n ), \cr
\rr_n
\,&\, \leq 3 \rr_{n-1}/4+  \rho_{n-1}^w/10
+\sigma^{n-9} |A_n|, \cr
\rho_n^w \,&\, \leq  \rho_{n-1}^w/10~.
}
$$
Thus, the
sequence of $A_n$ converges geometrically to a finite limit
$A_*$. Furthermore,
we find that
$ \lim_{n \rightarrow \infty} \rr_n = 0 $, and
$ \lim_{n \rightarrow \infty} \rho_n^w = 0 $.
Since the quantities
$ |A_n | $, $ \rr_n $, $  \rho_n^w $
increase only for at most 9 steps the term $ C R $
in \equ{ppsh} stays less than $ 1/2 $ if we choose
$ |A_1 | $, $ \rr_1 $, $  \rho_1^w  = \OO( \sigma^m ) $,
for a
sufficiently large $m>0$.
From \equ{ppsh} the existence of a finite constant
$ R = \sup_{n \in\natural } R_{\c\s,n}^u$ follows .
Finally,
the scaling of $ {w}_{n}(\cdot,\cdot,\tau)$ implies the exponential decay
of $ {w}(t)$.
The proof of \clm{main2} is complete.\QED

\def\bibitem[#1]#2#3#4\par{\endref
\ref\no{#1}\by{#3}\paper{\kern-0.2em#4}}
\def\sc{\eightpoint\rm}
\SECTIONNONR{References}
\raggedright\frenchspacing\raggedright
\eightpoint
\widestlabel{[EWW97]}
\ref
\bibitem[AW78]{AW78}
{\sc D.G. Aronson, H. Weinberger.}
Multidimensional nonlinear diffusion arising in population
genetics. {\it Adv. Math.} {\bf 30 }(1978), 33--76

\bibitem[BK92]{BK92}
{\sc J. Bricmont, A. Kupiainen.}
{Renormalization group and the Ginzburg--Landau equation.}
{\it Comm. Math. Phys.} {\bf 150 }(1992), 193--208

\bibitem[BK94]{BK94}
{\sc J. Bricmont, A. Kupiainen.}
{Stability of moving fronts in the Ginzburg--Landau equation.}
{\it Comm. Math. Phys.} {\bf 159 }(1994), 287--318

\bibitem[CE86]{CE86}
{\sc P. Collet, J.--P. Eckmann.}
The existence of dendritic fronts.
{\it Comm. Math. Phys.} {\bf 107 }(1986), 39--92

\bibitem[CE87]{CE87}
{\sc P. Collet, J.--P. Eckmann.}
{The stability of modulated fronts.}
{\it Helv. Phys. Acta} {\bf 60 }(1987), 969--991

\bibitem[CE90a]{CE90a}
{\sc P. Collet, J.--P. Eckmann.}
{\it Instabilities and fronts in extended systems.}
1990. Princeton, Princeton University Press

\bibitem[CE90b]{CE90b}
{\sc P. Collet, J.--P. Eckmann.}
{The time dependent amplitude equation for the Swift--Hohenberg problem.}
{\it Comm.\ Math.\ Phys.} {\bf 132} (1990), 139--153

\bibitem[CEE92]{CEE92}
{\sc P. Collet, J.--P. Eckmann, H. Epstein.}
{Diffusive repair for the   Ginsburg--Landau equation.}
{\it Helv. Phys. Acta} {\bf 65 }(1992), 56--92

\bibitem[DL83]{DL83}
{\sc G.\ Dee, J.\ S.\  Langer.}
Propagating pattern selection. {\it Phys. Rev. Lett.} {\bf 50 }(1983),
383--386

\bibitem[Eck65]{Eck65}
{\sc W. Eckhaus.}
{\it  Studies in nonlinear stability theory.}
Springer tracts in Nat. Phil. Vol. 6, 1965

\bibitem[EW91]{EW91}
{\sc J.--P. Eckmann, C.E. Wayne.}
{Propagating fronts and the center manifold theorem.}
{\it Comm. Math. Phys} {\bf 136 }(1991), 285--307

\bibitem[EW94]{EW94}
{\sc J.--P. Eckmann, C.E. Wayne.}
{The non--linear stability of front solutions for
parabolic partial differential equations.}
{\it Comm. Math. Phys.} {\bf 161 }(1994), 323--334

\bibitem[EWW97]{EWW97}
{\sc J.--P. Eckmann, C.E. Wayne, P. Wittwer.}
Geometric stability analysis of periodic solutions of the
Swift--Hohenberg equation.
{\it Comm. Math. Phys.} {\bf 190 }(1997), 173--211

\bibitem[Ga94]{Ga94}
{\sc T. Gallay.}
{Local stability of critical fronts in nonlinear parabolic
partial differential equations.}
{\it Nonlinearity} {\bf 7 }(1994), 741--764

\bibitem[HS99]{HS99}
{\sc M. Haragus, G. Schneider.}
Bifurcating fronts for the
Taylor--Couette problem
in infinite cylinders.
{\it Zeitschrift f\"ur Angewandte Mathematik und Physik (ZAMP)}
{\bf 50} (1999), 120--151

\bibitem[KSM92]{KSM92}
{\sc P. Kirrmann, G. Schneider, A. Mielke}
{The validity of modulation equations for extended systems with
cubic nonlinearities.} {\it Proceedings of the Royal Society of Edinburgh}
{\bf 122A} (1992), 85--91

\bibitem[RS72]{RS72}
{\sc M. Reed, B. Simon.}
{\it Methods of Modern Mathematical Physics I--IV.}
New York, Academic Press, 1972

\bibitem[Sa77]{Sa77}
{\sc D.H. Sattinger.}
{ Weighted norms for the stability
of travelling waves.}
{\it J. Diff. Eqns.} {\bf 25} (1977), 130--144

\bibitem[Schn94]{Schn94}
{\sc G. Schneider.}
{Error estimates for the Ginzburg--Landau approximation.}
{\it J. Appl. Math. Physics} {\bf 45} (1994), 433--457

\bibitem[Schn96]{Schn96}
{\sc G.\ Schneider.}
Diffusive stability of spatial periodic solutions
of the Swift--Hohenberg equation.
{\it Comm. Math. Phys.} {\bf 178 }(1996), 679--702

\bibitem[Schn98]{Schn98}
{\sc G.\ Schneider.}
Nonlinear stability of Taylor--vortices in
infinite cylinders.
{\it Archive for
Rational Mechanics and Analysis} {\bf 144 }(1998), 121--200

\bibitem[Ta97]{Ta97}
{\sc M.\ E.\ Taylor.}
Partial Differential Equations I: Basic Theory.Appl. Math. Sciences {\bf 115}, Springer 1997

\bibitem[vH91]{vH91}
{\sc A.\ van Harten.}
{On the validity of Ginzburg--Landau's equation.}
{\it J. Nonlinear Science} {\bf 1 }(1991), 397--422

\bibitem[Wa97]{Wa94}
{\sc C.E. Wayne.}
{Invariant manifolds for parabolic partial differential equations
on unbounded domains.}
{\it Arch.  Rat. Mech. Anal.} {\bf 138 }(1997), 279--306

\endref

\bye